\documentclass[aps,prd,twocolumn,superscriptaddress]{revtex4}
\usepackage{epsfig,epsf}
\usepackage{amsmath}
\usepackage{amsthm}
\usepackage{amsfonts}
\usepackage{amssymb}
\usepackage{dsfont}
\usepackage{multirow}
\usepackage{appendix}
\usepackage{slashed}
\usepackage[active]{srcltx}
\usepackage{psfrag}

\begin{document}

\title{Tensor states $\Upsilon B_{c}^{\ast -}$ and $J/\psi B_{c}^{\ast +}$}
\date{\today }
\author{S.~S.~Agaev}
\affiliation{Institute for Physical Problems, Baku State University, Az--1148 Baku,
Azerbaijan}
\author{K.~Azizi}
\affiliation{Department of Physics, University of Tehran, North Karegar Avenue, Tehran
14395-547, Iran}
\affiliation{Department of Physics, Dogus University, Dudullu-\"{U}mraniye, 34775
Istanbul, T\"{u}rkiye}
\author{H.~Sundu}
\affiliation{Department of Physics Engineering, Istanbul Medeniyet University, 34700
Istanbul, T\"{u}rkiye}

\begin{abstract}
Tensor states $\mathcal{M}_{\mathrm{T}}^{\mathrm{b}}=\Upsilon B_{c}^{\ast -}$
and $\mathcal{M}_{\mathrm{T}}^{\mathrm{c}}=J/\psi B_{c}^{\ast +}$ are
explored using techniques of QCD sum rule method. These hadronic molecules,
composed of only heavy quarks, have asymmetric quark contents $bb\overline{b}
\overline{c}$ and $cc\overline{c}\overline{b}$, respectively. The masses $%
m=(15864 \pm 85)~\mathrm{MeV}$ and $\widetilde{m}=(9870 \pm 82)~\mathrm{MeV}
$ prove that these structures are unstable against dissociations to
constituent mesons. Full widths of molecules $\mathcal{M}_{\mathrm{T}}^{
\mathrm{b}}$ and $\mathcal{M}_{\mathrm{T}}^{\mathrm{c}}$ are calculated by
considering their dominant and subleading decay channels. The subleading
channels are processes generated by annihilations of $\overline{b}b$ and $%
\overline{c}c$ quarks. For the molecule $\mathcal{M}_{\mathrm{T}}^{\mathrm{b}
}$ dominant decays are $\mathcal{M}_{\mathrm{T}}^{\mathrm{b}} \to \Upsilon
B_{c}^{\ast -}$ and $\mathcal{M}_{\mathrm{T}}^{\mathrm{b}} \to \eta_b
B_{c}^{-}$, whereas subleading channels are transformations to $\mathcal{M}%
_{ \mathrm{T}}^{\mathrm{b}}\rightarrow B^{(\ast )-}\overline{D}^{(\ast )0}$
and $\overline{B}_{(s)}^{(\ast )0}D_{(s)}^{(\ast )-}$ mesons. In the lower
limit ($\mathrm{l.l.}$) of the mass $m=15779~\mathrm{MeV}$ for $\mathcal{M}_{%
\mathrm{T}}^{\mathrm{b}}$ decay to $\Upsilon B_{c}^{\ast -}$ mesons is
forbidden. In the case of $\mathcal{M}_{\mathrm{T}}^{\mathrm{c}}$ we explore
decays to $J/\psi B_{c}^{\ast +}$, $\eta_{c}B_{c}^{+}$, $B^{(\ast)+}D^{(\ast
)0}$ and $B_{(s)}^{(\ast )0}D_{(s)}^{(\ast )+}$ mesons. Predictions $\Gamma[%
\mathcal{M} _{\mathrm{T}}^{\mathrm{b}}]=120^{+17}_{-12}~ \mathrm{MeV}$, $%
\Gamma[\mathcal{M}_{\mathrm{T}}^{\mathrm{b}}]_{\mathrm{l.l.}}=(65 \pm 7)~
\mathrm{MeV}$ and $\Gamma[ \mathcal{M}_{\mathrm{T}}^{\mathrm{c}}]=(71 \pm
9)~ \mathrm{MeV} $ for the widths of these molecules may be useful in
experimental studies of fully heavy structures.
\end{abstract}

\maketitle


\section{Introduction}

\label{sec:Intro}

The tetraquarks composed of four $b$ or/and $c$ quarks are intriguing
particles for high energy physics. Investigations of these unusual
structures may shed light on inner organization of both ordinary and exotic
hadrons \cite%
{Heller:1985cb,Lloyd:2003yc,Karliner:2016zzc,Anwar:2017:toa,Wu:2016vtq,Liu:2019zuc,Chen:2019vrj}%
. Numerous approaches, such as the chromomagnetic, the diquark, the
relativistic and nonrelativistic chiral quark and effective potential
models, were applied to study these exotic mesons. Their masses and widths
were calculated using the Bethe-Salpeter equations, QCD sum rule and other
methods \cite%
{Bedolla:2019zwg,Cordillo:2020sgc,Weng:2020jao,Wang:2021kfv,Deng:2020iqw,Yang:2021zrc, Galkin:2023wox,An:2022qpt,Wu:2024hrv,Hoffer:2024alv,Agaev:2024wvp,Agaev:2024mng,Agaev:2024qbh,Agaev:2024uza,Wang:2026gch}%
. There are various publications devoted to production mechanisms and
allowed decay channels of these exotic mesons \cite%
{Ali:2018ifm,Ali:2018xfq,Carvalho:2015nqf,Abreu:2023wwg}. Results of
performed explorations established the theoretical basis for new
investigations and provided valuable information for different collaborations

Many years theoretical analyses were only possible way to gain information
about fully heavy hadrons. Recent discoveries of LHCb-CMS-ATLAS groups
demonstrated that exotic mesons built of four heavy quarks, at least some of
them, are accessible in the current experiments \cite%
{LHCb:2020bwg,ATLAS:2023bft,CMS:2023owd,CMS:2026tiu}. In fact, the four $X$
structures observed by these collaborations placed the physics of fully
heavy tetraquarks on solid ground of experiments: These $X$ states are
supposedly exotic $cc\overline{c}\overline{c}$ tetraquarks.

The fully heavy tetraquarks with nonsymmetrical contents are also a class of
interesting states. In the diquark-antidiquark picture they were
investigated in numerous publications \cite%
{Galkin:2023wox,An:2022qpt,Wu:2024hrv,Wang:2026gch}. For instance, in Ref.\
\cite{Galkin:2023wox} the authors studied the mass spectra of such
diquark-antidiquark systems using the relativistic quark model and estimated
masses of tetraquarks with various quantum numbers. The tensor tetraquarks $%
bb\overline{b}\overline{c}$ and $cc\overline{c}\overline{b}$ in this model
have masses $16108~\mathrm{MeV}$ and $9620~\mathrm{MeV}$, respectively. The
similar analysis was done in Ref.\ \cite{An:2022qpt} by applying the
constituent quark model with results for the tensor tetraquarks $16149~%
\mathrm{MeV}$ and $9731~\mathrm{MeV}$. Interesting analyses were made in the
articles \cite{Wu:2024hrv,Wang:2026gch} as well.

In physics of all-heavy exotic mesons the diquark-antidiquark structure is
mostly employed model. Nevertheless, there are works in which authors
considered such systems in the framework of the molecule model. In this
picture four heavy quarks are grouped into colorless heavy mesons which form
the hadronic molecule. Thus, in Refs.\ \cite%
{Liu:2023gla,Agaev:2025wdj,Agaev:2025fwm,Agaev:2025nkw} this model was
utilized to explore properties of the molecular states $B_{c}^{(\ast )\pm
}B_{c}^{(\ast )\mp }$.

The hadronic molecules with asymmetric heavy quark contents were
investigated in Refs.\ \cite{Liu:2024pio,Agaev:2025did,Agaev:2025wyf}. The
extended local gauge formalism was used in Ref.\ \cite{Liu:2024pio} to
evaluate parameters of the asymmetric molecules with spin-parities $J^{%
\mathrm{P}}=0^{+}$, $1^{+}$. Calculations led to predictions that some of
the $bb\overline{b}\overline{c}$ molecules maybe are bound states.

The scalar $\eta _{b}B_{c}^{-}$, $\eta _{c}B_{c}^{-}$ and axial-vector $%
\Upsilon B_{c}^{-}$, $\eta _{b}B_{c}^{\ast -}$ and $J/\psi B_{c}^{+}$, $\eta
_{c}B_{c}^{\ast +}$ hadronic molecules were explored in our works \cite%
{Agaev:2025did,Agaev:2025wyf,Agaev:2025qgg}. The spectroscopic parameters of
these particles were evaluated in the context of the sum rule (SR) method
\cite{Shifman:1978bx,Shifman:1978by}. It turned out that, in the low-mass
scenario, the scalar molecule $\mathcal{M}_{\mathrm{b}}=\eta _{b}B_{c}^{-}$
with mass $15638~\mathrm{MeV}$ is below two-meson threshold $\eta
_{b}B_{c}^{-}$, hence $\mathcal{M}_{\mathrm{b}}$ is a bound state. The same
is true also in the case of the molecules $\Upsilon B_{c}^{-}$ and $\eta
_{b}B_{c}^{\ast -}$ which also may form bound states. But even in this case,
due to annihilation mechanism, these molecules are unstable against strong
decays to $B_{(s)}D_{(s)}$ meson pairs with correct quantum numbers and
charges.

In present paper, we extend our studies of the hadronic molecules with
nonsymmetrical heavy quark organizations, and, for the first time, calculate
the masses and decay widths of relevant tensor structures. We analyze the
molecules $\mathcal{M}_{\mathrm{T}}^{\mathrm{b}}=\Upsilon B_{c}^{\ast -}$
and $\mathcal{M}_{\mathrm{T}}^{\mathrm{c}}=J/\psi B_{c}^{\ast +}$ built of
heavy vector mesons. We are going to calculate their masses $m$, $\widetilde{%
m}$ and current couplings $\Lambda $, $\widetilde{\Lambda }$ by means of SR
method. The full decay widths of these molecules are evaluated as well. To
this end, we compute the partial widths of the leading and subleading decay
channels of $\mathcal{M}_{\mathrm{T}}^{\mathrm{b}}$ and $\mathcal{M}_{%
\mathrm{T}}^{\mathrm{c}}$. The leading decay modes of the molecule $\mathcal{%
M}_{\mathrm{T}}^{\mathrm{b}}$ are processes $\mathcal{M}_{\mathrm{T}}^{%
\mathrm{b}}\rightarrow \Upsilon B_{c}^{\ast -}$, and $\mathcal{M}_{\mathrm{T}%
}^{\mathrm{b}}\rightarrow \eta _{b}B_{c}^{-}$, whereas decays $\mathcal{M}_{%
\mathrm{T}}^{\mathrm{c}}\rightarrow J/\psi B_{c}^{\ast +}$ and $\mathcal{M}_{%
\mathrm{T}}^{\mathrm{c}}\rightarrow \eta _{c}B_{c}^{+}$ are dominant ones
for $\mathcal{M}_{\mathrm{T}}^{\mathrm{c}}$. In the leading modes of $%
\mathcal{M}_{\mathrm{T}}^{\mathrm{b}}$ and $\mathcal{M}_{\mathrm{T}}^{%
\mathrm{c}}$ all quarks (antiquarks) from these states appear in final-state
mesons. Second type of decays emerges after annihilations $b\overline{b}$ or
$c\overline{c}\rightarrow $ $q\overline{q}$, $s\overline{s}$ and subsequent
generation of $B_{(s)}^{(\ast )}D_{(s)}^{(\ast )}$ mesons \cite%
{Becchi:2020mjz,Becchi:2020uvq,Agaev:2023ara}. In the case of the molecule $%
\mathcal{M}_{\mathrm{T}}^{\mathrm{b}}$, we consider the six channels $%
\mathcal{M}_{\mathrm{T}}^{\mathrm{b}}\rightarrow B^{(\ast )-}\overline{D}%
^{(\ast )0},\ \overline{B}^{(\ast )0}D^{(\ast )-}$ and $\overline{B}%
_{s}^{(\ast )0}D_{s}^{(\ast )-}$. The subleading modes of $\mathcal{M}_{%
\mathrm{T}}^{\mathrm{c}}$ are decays to $B^{(\ast )+}D^{(\ast )0},\ B^{(\ast
)0}D^{(\ast )+}$ and $B_{s}^{(\ast )0}D_{s}^{(\ast )+}$ pairs. To explore
all these modes we make use of the three-point SR method which permits us to
estimate strong couplings at the molecule-meson-meson vertices of interest.

Presentation in this article is separated into six parts: In Sec.\ \ref%
{sec:Mass}, we compute the spectroscopic parameters of the molecules $%
\mathcal{M}_{\mathrm{T}}^{\mathrm{b}}$ and $\mathcal{M}_{\mathrm{T}}^{%
\mathrm{c}}$ . Processes $\mathcal{M}_{\mathrm{T}}^{\mathrm{b}}\rightarrow
\Upsilon B_{c}^{\ast -}$ and $\mathcal{M}_{\mathrm{T}}^{\mathrm{b}%
}\rightarrow \eta _{b}B_{c}^{-}$ are analyzed in Sec.\ \ref{sec:Widths1},
whereas the subleading channels of $\mathcal{M}_{\mathrm{T}}^{\mathrm{b}}$
are addressed in Sec.\ \ref{sec:Widths2}. Here, we estimate the full width
of $\mathcal{M}_{\mathrm{T}}^{\mathrm{b}}$ as well. The lower mass limit of
the molecule $\mathcal{M}_{\mathrm{T}}^{\mathrm{b}}$ is considered in this
section as well.\qquad\ The section \ref{sec:Widths3} is devoted to decays
of the hadronic molecule $\mathcal{M}_{\mathrm{T}}^{\mathrm{c}}$ and
calculation of its full width. We make our conclusions in the last part of
the paper\ \ref{sec:Conc}.


\section{The mass and current coupling of the molecules $\mathcal{M}_{%
\mathrm{T}}^{\mathrm{b}}$ and $\mathcal{M}_{\mathrm{T}}^{\mathrm{c}}$}

\label{sec:Mass}

The masses and current couplings of the hadronic molecules $\mathcal{M}_{%
\mathrm{T}}^{\mathrm{b}}$ and $\mathcal{M}_{\mathrm{T}}^{\mathrm{c}}$ are
among their important parameters. First, we concentrate on the mass $m$ and
current coupling $\Lambda $ of the state $\mathcal{M}_{\mathrm{T}}^{\mathrm{b%
}}$. These parameters can be extracted from the relevant sum rules. For
these purposes, we analyze the correlation function
\begin{equation}
\Pi _{\mu \nu \alpha \beta }(p)=i\int d^{4}xe^{ipx}\langle 0|\mathcal{T}%
\{I_{\mu \nu }(x)I_{\alpha \beta }^{\dag }(0)\}|0\rangle ,  \label{eq:CF1}
\end{equation}%
where $I_{\mu \nu }(x)$ is the interpolating current for the tensor molecule$%
\mathcal{M}_{\mathrm{T}}^{\mathrm{b}}$, and $\mathcal{T}$ \ is the
time-ordered product of two currents.

The current $I_{\mu \nu }(x)$ for the molecule $\mathcal{M}_{\mathrm{T}}^{%
\mathrm{b}}=\Upsilon B_{c}^{\ast -}$ has the form
\begin{equation}
I_{\mu \nu }(x)=\overline{b}_{a}(x)\gamma _{\mu }b_{a}(x)\overline{c}%
_{b}(x)\gamma _{\nu }b_{b}(x),  \label{eq:CR1}
\end{equation}%
and describes the state with spin-parity $J^{\mathrm{P}}=2^{+}$. Here, $a$
and $b$ are the the color indices.

The SRs for $m$ and $\Lambda $ is found by equating the correlators $\Pi
_{\mu \nu \alpha \beta }^{\mathrm{Phys}}(p)$ and $\Pi _{\mu \nu \alpha \beta
}^{\mathrm{OPE}}(p)$. The correlation function expressed by employing the
mass and current coupling of the molecule $\mathcal{M}_{\mathrm{T}}^{\mathrm{%
b}}$ gives $\Pi _{\mu \nu \alpha \beta }^{\mathrm{Phys}}(p)$. It is
determined by inserting into $\Pi _{\mu \nu \alpha \beta }(p)$ a complete
set of intermediate states, and performing a required integration over $x$.
Then $\Pi _{\mu \nu \alpha \beta }^{\mathrm{Phys}}(p)$ becomes equal to
\begin{eqnarray}
\Pi _{\mu \nu \alpha \beta }^{\mathrm{Phys}}(p) &=&\frac{\langle 0|I_{\mu
\nu }|\mathcal{M}_{\mathrm{T}}^{\mathrm{b}}(p,\epsilon )\rangle \langle
\mathcal{M}_{\mathrm{T}}^{\mathrm{b}}(p,\epsilon )|I_{\alpha \beta }^{\dag
}|0\rangle }{m^{2}-p^{2}}  \notag \\
&&+\cdots .  \label{eq:Phys1}
\end{eqnarray}%
The term written down above explicitly corresponds to the ground-level state
$\mathcal{M}_{\mathrm{T}}^{\mathrm{b}}$, while the dots stand for effects of
higher resonances and continuum states. We continue by introducing an
expression
\begin{equation}
\langle 0|I_{\mu \nu }|\mathcal{M}_{\mathrm{T}}^{\mathrm{b}}(p,\epsilon
(p))\rangle =\Lambda \epsilon _{\mu \nu }^{(\lambda )}(p),  \label{eq:ME1}
\end{equation}%
where $\epsilon =\epsilon _{\mu \nu }^{(\lambda )}(p)$ is the polarization
tensor of the molecule $\mathcal{M}_{\mathrm{T}}^{\mathrm{b}}$. We use Eq. (%
\ref{eq:ME1}) \ in the correlation function $\Pi _{\mu \nu \alpha \beta }^{%
\mathrm{Phys}}(p)$ and carry out necessary manipulations. These calculations
give
\begin{eqnarray}
\Pi _{\mu \nu \alpha \beta }^{\mathrm{Phys}}(p) &=&\frac{\Lambda ^{2}}{%
m^{2}-p^{2}}\left\{ \frac{1}{2}\left( g_{\mu \alpha }g_{\nu \beta }+g_{\mu
\beta }g_{\nu \alpha }\right) \right.  \notag \\
&&\left. +\text{ other terms}\right\} +\cdots .  \label{eq:Phys2}
\end{eqnarray}%
As is seen, $\Pi _{\mu \nu \alpha \beta }^{\mathrm{Phys}}(p)$ contains terms
with different Lorentz structures. The term $\sim (g_{\mu \alpha }g_{\nu
\beta }+g_{\mu \beta }g_{\nu \alpha })$ appears owing to a contribution of
the spin-$2$ particle. Remaining terms in $\Pi _{\mu \nu \alpha \beta }^{%
\mathrm{Phys}}(p)$ are admixtures of contributions arising from particles
with different spins. Therefore, it is convenient to employ $\sim (g_{\mu
\alpha }g_{\nu \beta }+g_{\mu \beta }g_{\nu \alpha })$ in our analysis and
denote corresponding invariant amplitude by $\Pi ^{\mathrm{Phys}}(p^{2})$.

To calculate $\Pi _{\mu \nu \alpha \beta }^{\mathrm{OPE}}(p)$ we substitute $%
I_{\mu \nu }(x)$ into $\Pi _{\mu \nu \alpha \beta }(p)$ and contracts quark
fields. As a result, we find
\begin{eqnarray}
&&\Pi _{\mu \nu \alpha \beta }^{\mathrm{OPE}}(p)=i\int d^{4}xe^{ipx}\left\{
\mathrm{Tr}\left[ \gamma _{\mu }S_{b}^{ab^{\prime }}(x)\gamma _{\beta
}S_{c}^{b^{\prime }b}(-x)\right. \right.  \notag \\
&&\left. \times \gamma _{\nu }S_{b}^{ba^{\prime }}(x)\gamma _{\alpha
}S_{b}^{a^{\prime }a}(-x)\right] -\mathrm{Tr}\left[ \gamma _{\mu
}S_{b}^{aa^{\prime }}(x)\gamma _{\alpha }S_{b}^{a^{\prime }a}(-x)\right]
\notag \\
&&\left. \times \mathrm{Tr}\left[ \gamma _{\nu }S_{b}^{bb^{\prime
}}(x)\gamma _{\beta }S_{c}^{b^{\prime }b}(-x)\right] \right\} ,
\label{eq:QCD1}
\end{eqnarray}%
where $S_{b(c)}(x)$ are propagators of heavy quarks \cite{Agaev:2020zad}.
The correlator $\Pi _{\mu \nu \alpha \beta }^{\mathrm{OPE}}(p)$ found after
these computations should be calculated with some accuracy using operator
product expansion ($\mathrm{OPE}$) .

After obtaining the term proportional to $(g_{\mu \alpha }g_{\nu \beta
}+g_{\mu \beta }g_{\nu \alpha })$ in $\Pi _{\mu \nu \alpha \beta }^{\mathrm{%
OPE}}(p)$ and denoting by $\Pi ^{\mathrm{OPE}}(p^{2})$ the corresponding
amplitude, we get the following SRs
\begin{equation}
m^{2}=\frac{\Pi ^{\prime }(M^{2},s_{0})}{\Pi (M^{2},s_{0})},  \label{eq:Mass}
\end{equation}%
and
\begin{equation}
\Lambda ^{2}=e^{m^{2}/M^{2}}\Pi (M^{2},s_{0}).  \label{eq:Coupl}
\end{equation}%
The function $\Pi (M^{2},s_{0})$ is the amplitude $\Pi ^{\mathrm{OPE}%
}(p^{2}) $ obtained after the Borel transformation and continuum
subtraction, whereas $\Pi ^{\prime }(M^{2},s_{0})$ is its derivative over $%
d/d(-1/M^{2})$. The Borel transformation is employed to suppress effects of
higher resonances and continuum states, while the subtraction procedure is
necessary to remove them from $\Pi (M^{2},\infty )$ in the context of the
quark-hadron duality assumption. The function $\Pi (M^{2},s_{0})$ depends on
$s_{0}$ and Borel parameter $M^{2}$. It is given by the expression
\begin{equation}
\Pi (M^{2},s_{0})=\int_{(3m_{b}+m_{c})^{2}}^{s_{0}}ds\rho ^{\mathrm{OPE}%
}(s)e^{-s/M^{2}}+\Pi (M^{2}).  \label{eq:CorrF}
\end{equation}%
The spectral density $\rho ^{\mathrm{OPE}}(s)$ is found as an imaginary part
of the function $\Pi ^{\mathrm{OPE}}(p^{2})$. It is calculated including
dimension-$4$ terms $\sim \langle \alpha _{s}G^{2}/\pi \rangle $. It is
worth noting that in fully-heavy systems where light quark and mixed
condensates are absent the triple-gluon condensate $\langle
g_{s}^{3}G^{3}\rangle $ constitutes the next nonperturbative contribution.
But in a situation when dimension-four effects are small contributions of
terms $\sim \langle g_{s}^{3}G^{3}\rangle $ can be safely neglected: This
fact was confirmed by explicit computations in Ref. \cite{Agaev:2025did}. In
Eq.\ (\ref{eq:CorrF}) $\rho ^{\mathrm{OPE}}(s)$ is the two-point spectral
density determined as the imaginary piece of the amplitude $\Pi ^{\mathrm{OPE%
}}(p^{2})$. The nonperturbative contribution $\Pi (M^{2})$ is evaluated
straightly from $\Pi ^{\mathrm{OPE}}(p^{2})$ and contains terms absent in $%
\rho ^{\mathrm{OPE}}(s)$.

For numerical calculations we have to fix some parameters in the relevant
SRs. The masses $m_{c}$ and $m_{b}$ and gluon condensate $\langle \alpha
_{s}G^{2}/\pi \rangle $ are well-known quantities. In the present article,
we employ $m_{c}=(1.2730\pm 0.0046)~\mathrm{GeV},$ $m_{b}=(4.183\pm 0.007)~%
\mathrm{GeV},$ and $\langle \alpha _{s}G^{2}/\pi \rangle =(0.012\pm 0.004)~%
\mathrm{GeV}^{4}.$The quark masses $m_{c}$ and $m_{b}$ are calculated in the
$\overline{\mathrm{MS}}$ scheme \cite{PDG:2024}, whereas gluon condensate $%
\langle \alpha _{s}G^{2}/\pi \rangle $ was estimated from studies of
numerous processes \cite{Shifman:1978bx,Shifman:1978by}.

The parameters $M^{2}$ and $s_{0}$ are specific for each analysis and have
to meet the standard constraints of the SR calculations. Thus, the pole
contribution ($\mathrm{PC}$) should dominate in the spectroscopic parameters
of the molecule $\mathcal{M}_{\mathrm{T}}^{\mathrm{b}}$ extracted from the
SRs. The convergence of $\mathrm{OPE}$ and maximal independence of $m$ and $%
\Lambda $ on parameters $M^{2}$ and $s_{0}$ are important for the reliable
SR studies. To meet these conditions, we impose on $M^{2}$ and $s_{0}$ some
constraints.Thus, the pole contribution
\begin{equation}
\mathrm{PC}=\frac{\Pi (M^{2},s_{0})}{\Pi (M^{2},\infty )},  \label{eq:PC}
\end{equation}%
should be $\mathrm{PC}\geq 0.5$ ensuring its prevalence in $m$ and $\Lambda $%
. Because apart from perturbative term $\Pi (M^{2},s_{0})$ contains the
dimension-$4$ contribution $\Pi ^{\mathrm{Dim4}}(M^{2},s_{0})$, we control
fulfillment of the condition $|\Pi ^{\mathrm{Dim4}}(M^{2},s_{0})|\leq
0.05|\Pi (M^{2},s_{0})|$, and by this way guarantee convergence of $\mathrm{%
OPE}$. The maximal value of $M^{2}$ is determined by $\mathrm{PC}$ while
convergence of $\mathrm{OPE}$ allows one to fix its minimum.

\begin{figure}[h]
\includegraphics[width=8.5cm]{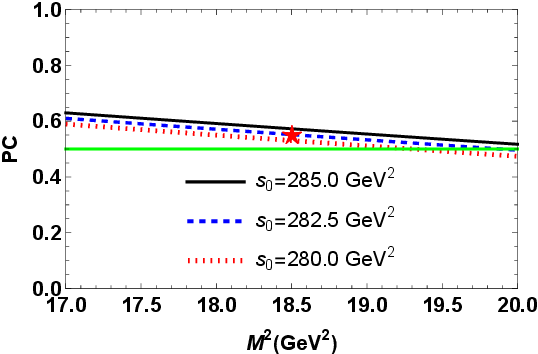}
\caption{Dependence of $\mathrm{PC}$ on $M^{2}$ for different $s_{0}$. The
constant line shows the border $\mathrm{PC}=0.5$. The star marks the point $%
M^{2}=18.5~\mathrm{GeV}^{2}$ and $s_{0}=282.5~\mathrm{GeV}^{2}$. }
\label{fig:PC}
\end{figure}

\begin{widetext}

\begin{figure}[htbp]
\begin{center}
\includegraphics[totalheight=6cm,width=8cm]{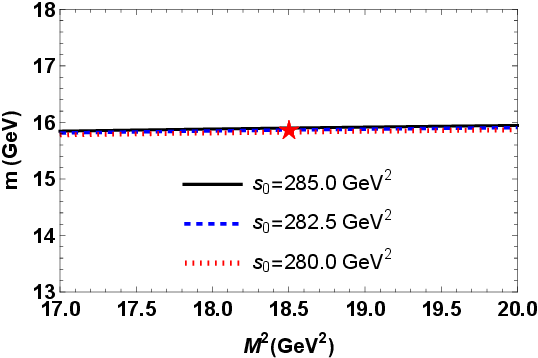}
\includegraphics[totalheight=6cm,width=8cm]{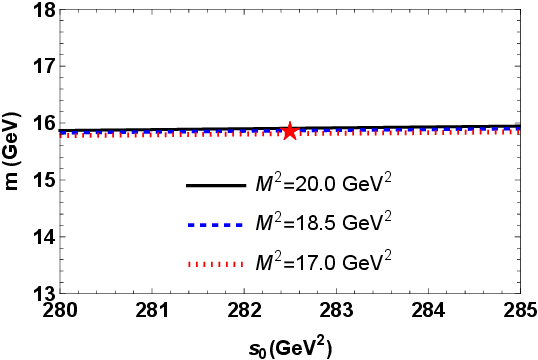}
\end{center}
\caption{Mass $m$ of $\mathcal{M}_{\mathrm{T}}^{\mathrm{b}}$ as a function of the Borel  $M^{2}$ (left), and continuum threshold $s_0$ parameters (right).}
\label{fig:Mass}
\end{figure}

\end{widetext}

Analyses are done at various values of $M^{2}$ and $s_{0}$, which allow us
to fix regions for $M^{2}$ and $s_{0}$, where all restrictions are obeyed.
We find that the windows
\begin{equation}
M^{2}\in \lbrack 17,20]~\mathrm{GeV}^{2},\ s_{0}\in \lbrack 280,285]~\mathrm{%
GeV}^{2}  \label{eq:Wind1}
\end{equation}%
satisfy aforementioned restrictions. Indeed, the pole contribution on the
average in $s_{0}$ is $\mathrm{PC}\approx 0.50$ and $\mathrm{PC}\approx 0.61$
at maximal and minimal values of $M^{2}$, respectively. The nonperturbative
term at $M^{2}=18.5~\mathrm{GeV}^{2}$ establishes less than $1.7\%$ of the
whole result. In Fig.\ \ref{fig:PC} we plot dependence of $\mathrm{PC}$ on
the Borel parameter $M^{2}$, where almost all lines exceed $\mathrm{PC}=0.5$
limit.

We evaluate $m$ and $\Lambda $ as their average values in the regions Eq.\ (%
\ref{eq:Wind1}) and get
\begin{eqnarray}
&&m=(15864\pm 85)~\mathrm{MeV},  \notag \\
&&\Lambda =(4.97\pm 0.45)~\mathrm{GeV}^{5}.  \label{eq:Result1}
\end{eqnarray}%
The results in Eq.\ (\ref{eq:Result1}) are effectively equal to SR
predictions at the point $M^{2}=18.5~\mathrm{GeV}^{2}$ and $s_{0}=282.5~%
\mathrm{GeV}^{2}$. In other words, this point is not chosen to evaluate
parameters of $\mathcal{M}_{\mathrm{T}}^{\mathrm{b}}$ but found in the
result of computations. At this point $\mathrm{PC}\approx 0.55$ which
ensures dominance of $\mathrm{PC}$ in the parameters $m$ and $\Lambda $.
Uncertainties in Eq.\ (\ref{eq:Result1}) are mainly due to ambiguities in $%
M^{2}$ and $s_{0}$: Errors connected with quark $m_{b(c)}$ and $\langle
\alpha _{s}G^{2}/\pi \rangle $ are very small. Errors in the mass $m$
constitutes $\pm 0.5\%$ of its value that proves reliability of the
extracted prediction. Ambiguities of the parameter $\Lambda $ equal to $\pm
9.1\%$ which is, at the same time, within limits usual for SR
investigations. In Fig.\ \ref{fig:Mass}, we plot $m$ as a function of $M^{2}$
and $s_{0}$.

With these results at hand, it is possible to check self-consistency of
performed analysis. Indeed, it is known that the continuum threshold
parameter $s_{0}$ separates the ground-level particle from its excited
states. Evidently, the mass $m$ of the molecule $\mathcal{M}_{\mathrm{T}}^{%
\mathrm{b}}$ should satisfy $m^{2}<s_{0}$. One can be easily convinced that
this constraint is met by parameters of the current calculations. Moreover, $%
s_{0}$ appears here also as a lower limit for the mass $m^{\ast 2}$ of the
first radially excited molecule $\mathcal{M}_{\mathrm{T}}^{\mathrm{b}}$
which has to be $s_{0}\leq m^{\ast 2}$. This allows one to fix the limit $m^{\ast }\geq 16.8~\mathrm{GeV}$. $\ $

\begin{widetext}

\begin{figure}[htbp]
\begin{center}
\includegraphics[totalheight=6cm,width=8cm]{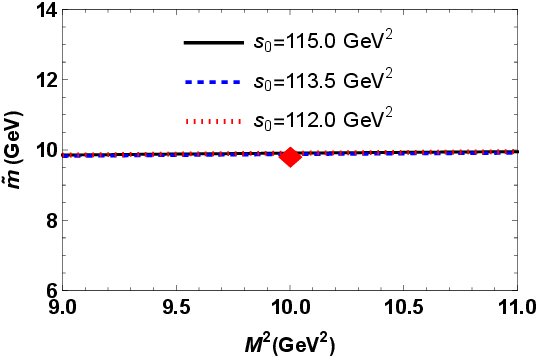}
\includegraphics[totalheight=6cm,width=8cm]{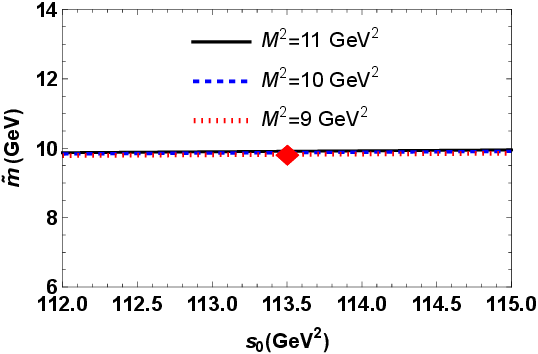}
\end{center}
\caption{Dependence of the mass $\widetilde{m}$ on the Borel  $M^{2}$ (left), and continuum threshold $s_0$ parameters (right).}
\label{fig:Mass1}
\end{figure}

\end{widetext}

The hadronic molecule $\mathcal{M}_{\mathrm{T}}^{\mathrm{c}}=J/\psi
B_{c}^{\ast +}$ is explored in accordance with the scheme explained above.
The interpolating current for $\mathcal{M}_{\mathrm{T}}^{\mathrm{c}}$ is
\begin{equation}
\widetilde{I}_{\mu \nu }(x)=\overline{c}_{a}(x)\gamma _{\mu }c_{a}(x)%
\overline{b}_{b}(x)\gamma _{\nu }c_{b}(x),  \label{eq:CR2}
\end{equation}%
We omit details and write down final predictions for parameters $\widetilde{m%
}$ and $\widetilde{\Lambda }$
\begin{eqnarray}
&&\widetilde{m}=(9870\pm 82)~\text{\textrm{M}}\mathrm{eV},  \notag \\
&&\widetilde{\Lambda }=(8.5\pm 0.9)\times 10^{-1}~\mathrm{GeV}^{5}.
\label{eq:Result2}
\end{eqnarray}%
Results in Eq.\ (\ref{eq:Result2}) have been obtained using for parameters $%
M^{2},\ s_{0}$ windows%
\begin{equation}
M^{2}\in \lbrack 9,11]~\mathrm{GeV}^{2},\ s_{0}\in \lbrack 112,115]~\mathrm{%
GeV}^{2}.  \label{eq:Wind1A}
\end{equation}%
The mass $\widetilde{m}$ of the molecule $\mathcal{M}_{\mathrm{T}}^{\mathrm{c%
}}$ is depicted in Fig.\ \ref{fig:Mass1} as a function of $M^{2}$ and $s_{0}
$.

\section{Leading decays $\mathcal{M}_{\mathrm{T}}^{\mathrm{b}}\rightarrow
\Upsilon B_{c}^{\ast -}$, $\protect\eta _{b}B_{c}^{-}$}

\label{sec:Widths1}


Mass of the molecule $\mathcal{M}_{\mathrm{T}}^{\mathrm{b}}$ fixes its
kinematically permitted decay modes. Actually there are two scenarios for
leading decay channels of the molecule $\mathcal{M}_{\mathrm{T}}^{\mathrm{b}}
$. In the first scenario, $\mathcal{M}_{\mathrm{T}}^{\mathrm{b}}$ has the
mass $m=15864~\mathrm{MeV}$ and decays to mesons $\Upsilon B_{c}^{\ast -}$
and $\eta _{b}B_{c}^{-}$ which are its leading (dominant)  modes. Indeed,
thresholds for these decays $15799~$\textrm{M}$\mathrm{eV}$ and $15674~%
\mathrm{MeV}$ do not overshoot the mass $m$. In the second scenario, i.e.,
in the lower limit for $m=15779~\mathrm{MeV}$ we see that the process $%
\mathcal{M}_{\mathrm{T}}^{\mathrm{b}}\rightarrow \Upsilon B_{c}^{\ast -}$
becomes kinematically forbidden for $\mathcal{M}_{\mathrm{T}}^{\mathrm{b}}$.
Nevertheless, the decay $\mathcal{M}_{\mathrm{T}}^{\mathrm{b}}\rightarrow
\eta _{b}B_{c}^{-}$ is still possible mode of $\mathcal{M}_{\mathrm{T}}^{%
\mathrm{b}}$.

\begin{figure}[h]
\includegraphics[width=8.5cm]{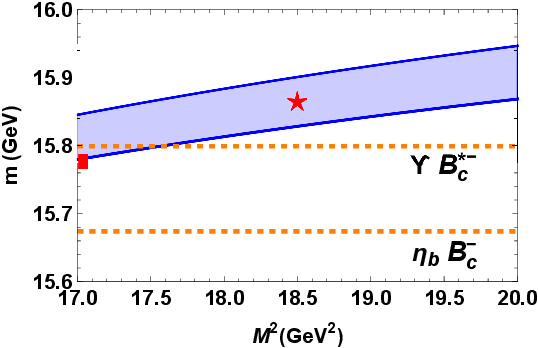}
\caption{The mass $m$ of the molecule $\mathcal{M}_{\mathrm{T}}^{\mathrm{b%
}}$ in the explored regions of parameters $M^{2}$ and $s_{0}$. The masses $%
m=15864~\mathrm{MeV}$ and $m=15779~\mathrm{MeV}$ are denoted by the star and
rectangle, respectively. The two-meson thresholds are shown by dashed lines.
}
\label{fig:Thresholds}
\end{figure}

These options are shown in Fig.\ \ref{fig:Thresholds}, in which one can be
convinced that almost in the whole region for the mass $m$ the molecule $%
\mathcal{M}_{\mathrm{T}}^{\mathrm{b}}$ is a resonant state and decays
through leading processes $\mathcal{M}_{\mathrm{T}}^{\mathrm{b}}\rightarrow
\Upsilon B_{c}^{\ast -}$, $\eta _{b}B_{c}^{-}$. Only in a small region of
the mass it transforms solely to mesons $\eta _{b}B_{c}^{-}$.


\subsection{Process $\mathcal{M}_{\mathrm{T}}^{\mathrm{b}}\rightarrow
\Upsilon B_{c}^{\ast -}$}


First, we explore the channel $\mathcal{M}_{\mathrm{T}}^{\mathrm{b}%
}\rightarrow \Upsilon B_{c}^{\ast -}$ and compute its partial width. To this
end, we should find the coupling $g_{1}$ at the vertex $\mathcal{M}_{\mathrm{%
T}}^{\mathrm{b}}\Upsilon B_{c}^{\ast -}$. This can be done by evaluating the
form factor $g_{1}(q^{2})$ at the mass shell $q^{2}=m_{B_{c}^{\ast }}^{2}$.

The sum rule for the form factor $g_{1}(q^{2})$ is derived from analysis of
the three-point correlation function
\begin{eqnarray}
\Pi _{\mu \nu \alpha \beta }(p,p^{\prime }) &=&i^{2}\int
d^{4}xd^{4}ye^{ip^{\prime }y}e^{-ipx}\langle 0|\mathcal{T}\{I_{\mu
}^{\Upsilon }(y)  \notag \\
&&\times I_{\nu }^{B_{c}^{\ast }}(0)I_{\alpha \beta }^{\dagger
}(x)\}|0\rangle ,  \label{eq:CF1a}
\end{eqnarray}%
where $I_{\mu }^{\Upsilon }(x)$ and $I_{\nu }^{B_{c}^{\ast }}(x)$ are the
interpolating currents of the vector mesons $\Upsilon $ and $B_{c}^{\ast -}$%
. These currents have the forms
\begin{equation}
I_{\mu }^{\Upsilon }(x)=\overline{b}_{i}(x)\gamma _{\mu }b_{i}(x),\ I_{\nu
}^{B_{c}^{\ast }}(x)=\overline{c}_{j}(x)\gamma _{\nu }b_{j}(x),
\end{equation}%
where $i$ and $j$ are the color indices.

The physical side of the sum rule $\Pi _{\mu \nu \alpha \beta }^{\mathrm{Phys%
}}(p,p^{\prime })$ is found by expressing Eq.\ (\ref{eq:CF1a}) in terms of $%
\mathcal{M}_{\mathrm{T}}^{\mathrm{b}}$, $\Upsilon $ and $B_{c}^{\ast -}$
particles' parameters. By including into analysis the ground-level
particles, we transform $\Pi _{\mu \nu \alpha \beta }(p,p^{\prime })$ and
obtain
\begin{eqnarray}
&&\Pi _{\mu \nu \alpha \beta }^{\mathrm{Phys}}(p,p^{\prime })=\frac{\langle
0|I_{\mu }^{\Upsilon }|\Upsilon (p^{\prime },\varepsilon _{1})\rangle }{%
p^{\prime 2}-m_{\Upsilon }^{2}}\frac{\langle 0|I_{\nu }^{B_{c}^{\ast
}}|B_{c}^{\ast -}(q,\varepsilon _{2})\rangle }{q^{2}-m_{B_{c}^{\ast }}^{2}}
\notag \\
&&\times \langle \Upsilon (p^{\prime },\varepsilon _{1})B_{c}^{\ast
-}(q,\varepsilon _{2})|\mathcal{M}_{\mathrm{T}}^{\mathrm{b}}(p,\epsilon
)\rangle \frac{\langle \mathcal{M}_{\mathrm{T}}^{\mathrm{b}}(p,\varepsilon
)|I_{\alpha \beta }^{\dagger }|0\rangle }{p^{2}-m^{2}}  \notag \\
&&+\cdots ,  \label{eq:TP1}
\end{eqnarray}%
where $m_{\Upsilon }=(9460.40\pm 0.10)~\mathrm{MeV}$ and $m_{B_{c}^{\ast
}}=6338~\mathrm{MeV}$ are masses of the $\Upsilon $ and $B_{c}^{\ast -}$
mesons \cite{PDG:2024,Godfrey:2004ya}, while $\varepsilon _{1}$ and $%
\varepsilon _{2}$ are their polarization vectors.

Equation (\ref{eq:TP1}) can be recast into a convenient form. To this end,
we employ the matrix elements%
\begin{eqnarray}
\langle 0|I_{\mu }^{\Upsilon }|\Upsilon (p^{\prime },\varepsilon
_{1})\rangle &=&f_{\Upsilon }m_{\Upsilon }\varepsilon _{1\mu }(p^{\prime }),
\notag \\
\langle 0|I_{\nu }^{B_{c}^{\ast }}|B_{c}^{\ast -}(q,\varepsilon _{2})\rangle
&=&f_{B_{c}^{\ast }}m_{B_{c}^{\ast }}\varepsilon _{2\nu }(q).  \label{eq:C2}
\end{eqnarray}%
Here, $f_{\Upsilon }=(708\pm 8)~\mathrm{MeV}$ and $f_{B_{c}^{\ast }}=471~%
\mathrm{MeV}$ are the mesons' decay constants\ \cite%
{Lakhina:2006vg,Eichten:2019gig}.

We should specify the matrix element $\langle \Upsilon (p^{\prime
},\varepsilon _{1})J/\psi (q,\varepsilon _{2})|\mathcal{M}_{\mathrm{T}%
}(p,\epsilon )\rangle $ which can be done by expressing it using the momenta
and polarization vectors of particles $\mathcal{M}_{\mathrm{T}}$, $\Upsilon $
and $J/\psi $ and corresponding form factors. Detailed analysis confirms
that the tensor-vector-vector vertex has the form \cite{Agaev:2024pil}
\begin{eqnarray}
&&\langle \Upsilon (p^{\prime },\varepsilon _{1})B_{c}^{\ast
-}(q,\varepsilon _{2})|\mathcal{M}_{\mathrm{T}}^{\mathrm{b}}(p,\epsilon
)\rangle =g_{1}(q^{2})\epsilon _{\tau \rho }^{(\lambda )}\left[ (\varepsilon
_{1}^{\ast }\cdot q)\right.  \notag \\
&&\left. \times \varepsilon _{2}^{\tau \ast }p^{\prime \rho }+(\varepsilon
_{2}^{\ast }\cdot p^{\prime })\varepsilon _{1}^{\ast \tau }q^{\rho
}-(p^{\prime }\cdot q)\varepsilon _{1}^{\tau \ast }\varepsilon _{2}^{\rho
\ast }-(\varepsilon _{1}^{\ast }\cdot \varepsilon _{2}^{\ast })p^{\prime
\tau }q^{\rho }\right] .  \notag \\
&&  \label{eq:TVV}
\end{eqnarray}%
Then, for $\Pi _{\mu \nu \alpha \beta }^{\mathrm{Phys}}(p,p^{\prime })$ we
get
\begin{eqnarray}
&&\Pi _{\mu \nu \alpha \beta }^{\mathrm{Phys}}(p,p^{\prime })=g_{1}(q^{2})%
\frac{\Lambda f_{\Upsilon }m_{\Upsilon }f_{J/\psi }m_{J/\psi }}{\left(
p^{2}-m^{2}\right) (p^{\prime 2}-m_{\Upsilon }^{2})(q^{2}-m_{B_{c}^{\ast
}}^{2})}  \notag \\
&&\times \left[ p_{\beta }^{\prime }p_{\alpha }^{\prime }g_{\mu \nu }+\frac{1%
}{2}p_{\mu }p_{\alpha }^{\prime }g_{\beta \nu }+\frac{1}{2m^{2}}p_{\beta
}p_{\nu }p_{\mu }^{\prime }p_{\alpha }^{\prime }\right.  \notag \\
&&\left. +\text{ other structures}\right] +\cdots .
\end{eqnarray}

The correlation function $\Pi _{\mu \nu \alpha \beta }^{\mathrm{OPE}%
}(p,p^{\prime })$ is given by the formula
\begin{eqnarray}
&&\Pi _{\mu \nu \alpha \beta }^{\mathrm{OPE}}(p,p^{\prime })=\int
d^{4}xd^{4}ye^{ip^{\prime }y}e^{-ipx}\left \lbrace\mathrm{Tr}\left[ \gamma _{\mu
}S_{b}^{ia}(y-x)\right.\right.  \notag \\
&&\left. \times \gamma _{\alpha }S_{b}^{ai}(x-y)\right] \mathrm{Tr}\left[
\gamma _{\nu }S_{b}^{jb}(-x)\gamma _{\beta }S_{b}^{bj}(x)\right]  \notag \\
&&-\left.\mathrm{Tr}\left[ \gamma _{\mu }S_{b}^{ib}(y-x)\gamma _{\beta
}S_{c}^{bj}(x)\gamma _{\nu }S_{b}^{ja}(-x)\gamma _{\alpha }S_{b}^{ai}(x-y)
\right] \right\rbrace .  \notag \\
&&  \label{eq:CF3}
\end{eqnarray}%
We use the amplitudes $\Pi _{1}^{\mathrm{Phys}}(p^{2},p^{\prime 2},q^{2})$
and $\Pi _{1}^{\mathrm{OPE}}(p^{2},p^{\prime 2},q^{2})$ that correspond to
terms $\sim p_{\beta }p_{\nu }p_{\mu }^{\prime }p_{\alpha }^{\prime }$ in
these correlators, and derive SR for $g_{1}(q^{2})$. After standard
operations, we find%
\begin{equation}
g_{1}(q^{2})=\frac{2m^{2}(q^{2}-m_{B_{c}^{\ast }}^{2})}{\Lambda f_{\Upsilon
}m_{\Upsilon }f_{B_{c}^{\ast }}m_{B_{c}^{\ast }}}e^{m^{2}/M_{1}^{2}}e^{m_{%
\Upsilon }^{2}/M_{2}^{2}}\Pi _{1}(\mathbf{M}^{2},\mathbf{s}_{0},q^{2}).
\label{eq:SRG}
\end{equation}%
In Eq.\ (\ref{eq:SRG}), $\Pi _{1}(\mathbf{M}^{2},\mathbf{s}_{0},q^{2})$ is
the function $\Pi _{1}^{\mathrm{OPE}}(p^{2},p^{\prime 2},q^{2})$ after the
Borel transformations and continuum subtractions. As is clear it contains
parameters $\mathbf{M}^{2}=(M_{1}^{2},M_{2}^{2})$ and $\mathbf{s}%
_{0}=(s_{0},s_{0}^{\prime })$ where the pairs $(M_{1}^{2},s_{0})$ and $%
(M_{2}^{2},s_{0}^{\prime })$ correspond to the channels of the molecule $%
\mathcal{M}_{\mathrm{T}}^{\mathrm{b}}$ and meson $\Upsilon $ , respectively.
It is determined in the following form%
\begin{eqnarray}
&&\Pi _{1}(\mathbf{M}^{2},\mathbf{s}_{0},q^{2})=%
\int_{(3m_{b}+m_{c})^{2}}^{s_{0}}ds\int_{4m_{b}^{2}}^{s_{0}^{\prime
}}ds^{\prime }\rho _{1}(s,s^{\prime },q^{2})  \notag \\
&&\times e^{-s/M_{1}^{2}-s^{\prime }/M_{2}^{2}}.  \label{eq:CorrF1}
\end{eqnarray}

Restrictions applied to find parameters $\mathbf{M}^{2}$ and $\mathbf{s}_{0}$
are usual for all sum rule studies and have been explained in Sec.\ \ref%
{sec:Mass}. Our computations demonstrate that windows for the parameters $%
(M_{1}^{2},s_{0})$ Eq.\ (\ref{eq:Wind1}) and
\begin{equation}
M_{2}^{2}\in \lbrack 9,11]~\mathrm{GeV}^{2},\ s_{0}^{\prime }\in \lbrack
98,100]~\mathrm{GeV}^{2}.  \label{eq:Wind3}
\end{equation}%
for $(M_{2}^{2},s_{0}^{\prime })$ obey these restrictions. Note that $%
s_{0}^{\prime }$ is bounded by the mass $m_{\Upsilon (2S)}=(10023.4\pm 0.5)~%
\mathrm{MeV}\ $of the meson $\Upsilon (2S)$, i.e., $s_{0}^{\prime
}<m_{\Upsilon (2S)}^{2}$.

\begin{figure}[h]
\includegraphics[width=8.5cm]{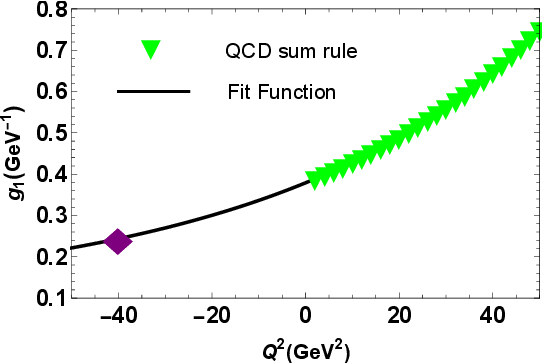}
\caption{QCD data and fit function for $g_{1}(Q^{2})$. The diamond shows the
point $Q^{2}=-m_{B_{c}^{\ast}}^{2}$ where $g_{1}$ has been evaluated. }
\label{fig:Fit}
\end{figure}

The sum rule for the form factor $g_{1}(q^{2})$ is applicable in the region $%
q^{2}<0$. But $g_{1}(q^{2})$ determines the coupling $g_{1}$ at the mass
shell $q^{2}=m_{B_{c}^{\ast }}^{2}$. For that reason, we introduce the
function $g_{1}(Q^{2})$ where $Q^{2}=-q^{2}$ and utilize it in our studies.
The SR results for $g_{1}(Q^{2})$ are demonstrated in Fig.\ \ref{fig:Fit},
where $Q^{2}$ changes in the interval $Q^{2}=2-50~\mathrm{GeV}^{2}$.

Above it has been noted that $g_{1}$ should be extracted at $%
q^{2}=m_{B_{c}^{\ast }}^{2}$, i.e., at $Q^{2}=-m_{B_{c}^{\ast }}^{2}$. But
at that point one can not use the SR method directly. To avoid this problem,
we employ the function $\mathcal{Z}_{1}(Q^{2})$ which at $Q^{2}>0$ amounts
to the SR data $g_{1}(Q^{2})$, but can be extrapolated to the region $%
Q^{2}<0 $. For these purposes, we employ
\begin{equation}
\mathcal{Z}_{i}(Q^{2})=\mathcal{Z}_{i}^{0}\mathrm{\exp }\left[ z_{i}^{1}%
\frac{Q^{2}}{m^{2}}+z_{i}^{2}\left( \frac{Q^{2}}{m^{2}}\right) ^{2}\right] ,
\label{eq:FitF}
\end{equation}%
where $\mathcal{Z}_{i}^{0}$, $z_{i}^{1}$, and $z_{i}^{2}$ are fitted
constants. From confronting of the SR data and Eq.\ (\ref{eq:FitF}), it is
easy to fix
\begin{equation}
\mathcal{Z}_{1}^{0}=0.379~\mathrm{GeV}^{-1},z_{1}^{1}=3.072,\text{ }%
z_{1}^{2}=1.850.  \label{eq:FF1}
\end{equation}%
The function $\mathcal{Z}_{1}(Q^{2})$ is drawn in Fig.\ \ref{fig:Fit}, where
agreement with the SR data is evident. For $g_{1}$, one gets
\begin{equation}
g_{1}\equiv \mathcal{Z}_{1}(-m_{B_{c}^{\ast }}^{2})=(2.43\pm 0.46)\times
10^{-1}\ \mathrm{GeV}^{-1}.  \label{eq:g1}
\end{equation}

Partial width of the decay $\mathcal{M}_{\mathrm{T}}^{\mathrm{b}}\rightarrow
\Upsilon B_{c}^{\ast -}$ is given by the formula%
\begin{equation}
\Gamma \left[ \mathcal{M}_{\mathrm{T}}^{\mathrm{b}}\rightarrow \Upsilon
B_{c}^{\ast -}\right] =g_{1}^{2}\frac{\lambda _{1}}{40\pi m^{2}}|M_{1}|^{2},
\label{eq:PDw2}
\end{equation}%
where%
\begin{eqnarray}
&&|M_{1}|^{2}=\frac{1}{6m^{4}}\left[ m_{B_{c}^{\ast }}^{8}+m_{B_{c}^{\ast
}}^{6}(m^{2}-4m_{\Upsilon }^{2})+(m^{2}-m_{\Upsilon }^{2})^{2}\right.  \notag
\\
&&\times (6m^{4}+3m^{2}m_{\Upsilon }^{2}+m_{\Upsilon }^{4})+m_{B_{c}^{\ast
}}^{4}(m^{4}-m^{2}m_{\Upsilon }^{2}+6m_{\Upsilon }^{4})  \notag \\
&&\left. -m_{B_{c}^{\ast }}^{2}(9m^{6}-34m^{4}m_{\Upsilon
}^{2}+m^{2}m_{\Upsilon }^{4}+4m_{\Upsilon }^{6})\right] .  \label{eq:M1}
\end{eqnarray}%
In Eq.\ (\ref{eq:PDw2}), $\lambda _{1}=\lambda (m,m_{\Upsilon
},m_{B_{c}^{\ast }})$ is defined as
\begin{equation}
\lambda (x,y,z)=\frac{\sqrt{%
x^{4}+y^{4}+z^{4}-2(x^{2}y^{2}+x^{2}z^{2}+y^{2}z^{2})}}{2x}.
\end{equation}%
Finally, one gets
\begin{equation}
\Gamma \left[ \mathcal{M}_{\mathrm{T}}^{\mathrm{b}}\rightarrow \Upsilon
B_{c}^{\ast -}\right] =24.2_{-6.5}^{+14.1}~\mathrm{MeV}.  \label{eq:DW2}
\end{equation}

The errors above arise owing to ambiguities of $g_{1}$ and masses of
particles $\Upsilon $, $B_{c}^{\ast -}$ and $\mathcal{M}_{\mathrm{T}}^{%
\mathrm{b}}$ (upper limit) in Eq.\ (\ref{eq:PDw2}).

\subsection{Decay $\mathcal{M}_{\mathrm{T}}^{\mathrm{b}}\rightarrow $ $%
\protect\eta _{b}B_{c}^{-}$}


Partial width of this process is determined by the coupling $g_{2}$ at the
vertex $\mathcal{M}_{\mathrm{T}}^{\mathrm{b}}\eta _{b}B_{c}^{-}$. To obtain
corresponding form factor $g_{2}(q^{2})$, one has to consider the correlator%
\begin{eqnarray}
\Pi _{\mu \nu }(p,p^{\prime }) &=&i^{2}\int d^{4}xd^{4}ye^{ip^{\prime
}y}e^{-ipx}\langle 0|\mathcal{T}\{\ I^{\eta _{b}}(y)  \notag \\
&&\times I^{B_{c}}(0)I_{\mu \nu }^{\dagger }(x)\}|0\rangle .  \label{eq:CF7}
\end{eqnarray}%
The interpolating currents of the mesons $\eta _{b}$ and $B_{c}^{-}$ are
\begin{equation}
\ I^{\eta _{b}}(x)=\overline{b}_{j}(x)i\gamma _{5}b_{j}(x),\ I^{B_{c}}(x)=%
\overline{c}_{i}(x)i\gamma _{5}b_{i}(x),  \label{eq:C3}
\end{equation}

To calculate $\Pi _{\mu \nu }^{\mathrm{Phys}}(p,p^{\prime })$ we use the
matrix elements
\begin{eqnarray}
&&\langle 0|I^{\eta _{b}}|\eta _{b}(p^{\prime })\rangle =\frac{f_{\eta
_{b}}m_{\eta _{b}}^{2}}{2m_{b}},  \notag \\
&&\langle 0|I^{B_{c}}|B_{c}^{-}(q)\rangle =\frac{f_{B_{c}}m_{B_{c}}^{2}}{%
m_{b}+m_{c}}.  \label{eq:ME4}
\end{eqnarray}%
Here, $f_{\eta _{b}}$, $m_{\eta _{b}}$ and $f_{B_{c}}$, $m_{B_{c}}$ are the
decay constants and masses of the mesons $\eta _{b}$ and $B_{c}^{-}$,
respectively. The vertex $\mathcal{M}_{\mathrm{T}}^{\mathrm{b}}\eta
_{b}B_{c}^{-}$ is determined by the expression \cite{Agaev:2024pil}%
\begin{equation}
\langle \eta _{b}(p^{\prime })B_{c}^{-}(q)|\mathcal{M}_{\mathrm{T}}^{\mathrm{%
b}}(p,\epsilon )\rangle =g_{2}(q^{2})\epsilon _{\alpha \beta }^{(\lambda
)}(p)p^{\prime \alpha }p^{\prime \beta }.
\end{equation}%
Then $\Pi _{\mu \nu }^{\mathrm{Phys}}(p,p^{\prime })$ is
\begin{eqnarray}
&&\Pi _{\mu \nu }^{\mathrm{Phys}}(p,p^{\prime })=)\frac{g_{2}(q^{2})\Lambda
f_{\eta _{b}}m_{\eta _{b}}^{2}f_{B_{c}}m_{B_{c}}^{2}}{2m_{b}(m_{b}+m_{c})%
\left( p^{2}-m^{2}\right) (p^{\prime 2}-m_{\eta _{b}}^{2})}  \notag \\
&&\times \frac{1}{(q^{2}-m_{B_{c}}^{2})}\left\{ \frac{1}{12m^{2}}\left[
m^{4}-2m^{2}(m_{\eta _{b}}^{2}+q^{2})\right. \right.  \notag \\
&&\left. \left. +(m_{\eta _{b}}^{2}-q^{2})^{2}\right] g_{\mu \nu }+p_{\mu
}^{\prime }p_{\nu }^{\prime }+\text{other terms}\right\} .  \label{eq:CF4}
\end{eqnarray}%
The function $\Pi _{\mu \nu }^{\mathrm{OPE}}(p,p^{\prime })$ reads
\begin{eqnarray}
&&\Pi _{\mu \nu }^{\mathrm{OPE}}(p,p^{\prime })=-\int
d^{4}xd^{4}ye^{ip^{\prime }y}e^{-ipx}\mathrm{Tr}\left[ \gamma
_{5}S_{b}^{ib}(y-x)\right.  \notag \\
&&\left. \times \gamma _{\nu }S_{c}^{bj}(x)\gamma _{5}S_{b}^{ja}(-x)\gamma
_{\mu }S_{b}^{ai}(x-y)\right] .  \label{eq:CF5}
\end{eqnarray}%
The terms in $\Pi _{\mu \nu }^{\mathrm{Phys}}(p,p^{\prime })$ and $\Pi _{\mu
\nu }^{\mathrm{OPE}}(p,p^{\prime })$ have the identical Lorentz structures.
The sum rule for $g_{2}(q^{2})$ is found using terms$\sim p_{\mu }^{\prime
}p_{\nu }^{\prime }$ and corresponding amplitudes $\Pi _{2}^{\mathrm{Phys}%
}(p^{2},p^{\prime 2},q^{2})$ and $\Pi _{2}^{\mathrm{OPE}}(p^{2},p^{\prime
2},q^{2})$. As a result, we get

\begin{eqnarray}
&&g_{2}(q^{2})=\frac{2m_{b}(m_{b}+m_{c})(q^{2}-m_{B_{c}}^{2})}{\Lambda
f_{B_{c}}m_{B_{c}}^{2}f_{\eta _{b}}m_{\eta _{b}}^{2}}  \notag \\
&&\times e^{m^{2}/M_{1}^{2}}e^{m_{\eta _{b}}^{2}/M_{2}^{2}}\Pi _{2}(\mathbf{M%
}^{2},\mathbf{s}_{0},q^{2}),  \label{eq:SRCoup}
\end{eqnarray}%
where $\Pi _{2}(\mathbf{M}^{2},\mathbf{s}_{0},q^{2})$ is the amplitude $\Pi
_{2}^{\mathrm{OPE}}(p^{2},p^{\prime 2},q^{2})$ after relevant
transformations and subtractions.

Operations to extract $g_{2}(q^{2})$ have been explained above therefore we
omit further details. In computations, we use the $B_{c}^{-}$ and $\eta _{b}$
mesons' masses i.e., $m_{B_{c}}=(6274.47\pm 0.27\pm 0.17)~\mathrm{MeV}$, $%
m_{\eta _{b}}=(9398.7~\pm 2.0)\ \mathrm{MeV}$ \cite{PDG:2024}. We use $%
f_{B_{c}}=(371\pm 37)~\mathrm{MeV}$, and $f_{\eta _{b}}=724~\mathrm{MeV}$ as
their decay constants, where $f_{B_{c}}$ is the result of SR analysis \cite%
{Wang:2024fwc}. In the $\eta _{b}$ meson channel we apply the working
windows
\begin{equation}
M_{2}^{2}\in \lbrack 9,11]~\mathrm{GeV}^{2},\ s_{0}^{\prime }\in \lbrack
95,99]~\mathrm{GeV}^{2}.
\end{equation}

The extrapolating function $\mathcal{Z}_{2}(Q^{2})$ with parameters $%
\mathcal{Z}_{2}^{0}=31.73~\mathrm{GeV}^{-1}$, $z_{2}^{1}=2.36$, and $%
z_{2}^{2}=3.88$ nicely agrees with QCD predictions. The strong coupling $%
g_{2}$ amounts to
\begin{equation}
g_{2}\equiv \mathcal{Z}_{2}(-m_{B_{c}}^{2})=(24.10\pm 4.34)\ \mathrm{GeV}%
^{-1}.
\end{equation}%
The partial width of this process is equal to
\begin{equation}
\Gamma \left[ \mathcal{M}_{\mathrm{T}}^{\mathrm{b}}\rightarrow \eta
_{b}B_{c}^{-}\right] =g_{2}^{2}\frac{\lambda _{2}}{40\pi m^{2}}|M_{2}|^{2},
\end{equation}%
where%
\begin{eqnarray}
&&|M_{2}|^{2}=\frac{\left[ m^{4}+(m_{\eta
_{b}}^{2}-m_{B_{c}}^{2})^{2}-2m^{2}(m_{\eta _{b}}^{2}+m_{B_{c}}^{2})\right]
^{2}}{24m^{4}},  \notag \\
&&  \label{eq:M2}
\end{eqnarray}%
and $\lambda _{2}=\lambda (m,m_{\eta _{b}},m_{B_{c}})$.

The decay $\mathcal{M}_{\mathrm{T}}^{\mathrm{b}}\rightarrow \eta
_{b}B_{c}^{-}$ has the width
\begin{equation}
\Gamma \left[ \mathcal{M}_{\mathrm{T}}^{\mathrm{b}}\rightarrow \eta
_{b}B_{c}^{-}\right] =(30.7\pm 6.2)~\mathrm{MeV}.  \label{eq:DW20}
\end{equation}


\section{Subleading channels of $\mathcal{M}_{\mathrm{T}}^{\mathrm{b}}$}

\label{sec:Widths2}


The hadronic molecule $\mathcal{M}_{\mathrm{T}}^{\mathrm{b}}$ can transform
to conventional mesons after annihilation of $b\overline{b}$ quarks to light
quark-antiquark pairs \cite{Becchi:2020mjz,Becchi:2020uvq,Agaev:2023ara} and
generation of $BD$ mesons with required spin-parities. We study here the
decays to $B^{-}\overline{D}^{0}$,$\ B^{\ast -}\overline{D}^{\ast 0}$, $%
\overline{B}^{0}D^{-}$, $\overline{B}^{\ast 0}D^{\ast -}$, $\overline{B}%
_{s}^{0}D_{s}^{-}$ and $\overline{B}_{s}^{\ast 0}D_{s}^{\ast -}$ meson pairs.

It should be noted that the correlation functions of decays, for example, $%
\mathcal{M}_{\mathrm{T}}^{\mathrm{b}}\rightarrow B^{-}\overline{D}^{0}$ and $%
\mathcal{M}_{\mathrm{T}}^{\mathrm{b}}\rightarrow \overline{B}^{0}D^{-}$
differ only by propagators of $u$ and $d$ quarks. Here we adopt the
approximation $m_{u}=m_{d}=0$, and also neglect small numerical differences
in the masses of charged and neutral $B$ and $D$ mesons, therefore processes
$\mathcal{M}_{\mathrm{T}}^{\mathrm{b}}\rightarrow B^{-}\overline{D}^{0}$ and
$\mathcal{M}_{\mathrm{T}}^{\mathrm{b}}\rightarrow \overline{B}^{0}D^{-}$
have the same decay widths.


\subsection{Decays $\mathcal{M}_{\mathrm{T}}^{\mathrm{b}}\rightarrow B^{\ast
-}\overline{D}^{\ast 0}$ and $\mathcal{M}_{\mathrm{T}}^{\mathrm{b}%
}\rightarrow \overline{B}^{\ast 0}D^{\ast -}$}


Let us consider the channel $\mathcal{M}_{\mathrm{T}}^{\mathrm{b}%
}\rightarrow B^{\ast -}\overline{D}^{\ast 0}$. To find the molecule-meson
meson coupling $g_{3}$ at the vertex $\mathcal{M}_{\mathrm{T}}^{\mathrm{b}%
}B^{\ast -}\overline{D}^{\ast 0}$, we explore the correlation function%
\begin{eqnarray}
\widehat{\Pi }_{\mu \nu \alpha \beta }(p,p^{\prime }) &=&i^{2}\int
d^{4}xd^{4}ye^{ip^{\prime }y}e^{-ipx}\langle 0|\mathcal{T}\{I_{\mu
}^{B^{\ast }}(y)  \notag \\
&&\times I_{\nu }^{\overline{D}^{\ast 0}}(0)I_{\alpha \beta }^{\dagger
}(x)\}|0\rangle ,  \label{eq:CF1A}
\end{eqnarray}%
with $I_{\mu }^{B^{\ast }}(x)$ and $I_{\nu }^{\overline{D}^{\ast 0}}(x)$
being the interpolating currents for the mesons $B^{\ast -}$ and $\overline{D%
}^{\ast 0}$
\begin{equation}
I_{\mu }^{B^{\ast }}(x)=\overline{u}_{j}(x)\gamma _{\mu }b_{j}(x),\ I_{\nu
}^{\overline{D}^{\ast 0}}(x)=\overline{c}_{i}(x)\gamma _{\nu }u_{i}(x),\text{
}  \label{eq:CRB}
\end{equation}

\begin{figure}[h]
\includegraphics[width=8.5cm]{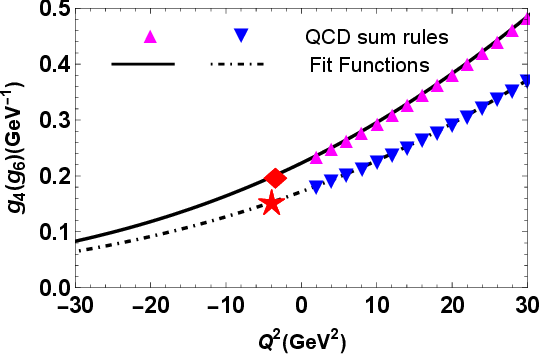}
\caption{SR data and fit functions for the form factors $g_{4}(Q^{2})$
(solid line) and $g_{6}(Q^{2})$ (dot-dashed line). The diamond and star mark
the points $Q^{2}=-m_{D^{0}}^{2}$ and $Q^{2}=-m_{D_{s}^{0}}^{2}$,
respectively. }
\label{fig:Fit1}
\end{figure}

The correlation function $\widehat{\Pi }_{\mu \nu \alpha \beta }(p,p^{\prime
})$ in terms of the matrix elements of the particles $\mathcal{M}_{\mathrm{T}%
}^{\mathrm{b}}$, $B^{\ast -}$, and $\overline{D}^{\ast 0}$ is
\begin{eqnarray}
&&\widehat{\Pi }_{\mu \nu \alpha \beta }^{\mathrm{Phys}}(p,p^{\prime })=%
\frac{\langle 0|I_{\mu }^{B^{\ast }}|B^{\ast -}(p^{\prime },\varepsilon
_{1})\rangle }{p^{\prime 2}-m_{B^{\ast }}^{2}}\frac{\langle 0|I_{\nu }^{%
\overline{D}^{\ast 0}}|\overline{D}^{\ast 0}(q,\varepsilon _{2})\rangle }{%
q^{2}-m_{D^{\ast 0}}^{2}}  \notag  \label{eq:CF2} \\
&&\times \langle B^{\ast -}(p^{\prime },\varepsilon _{1})\overline{D}^{\ast
0}(q,\varepsilon _{2})|\mathcal{M}_{\mathrm{T}}^{\mathrm{b}}(p,\epsilon
)\rangle \frac{\langle \mathcal{M}_{\mathrm{T}}^{\mathrm{b}}(p,\epsilon
)|I_{\alpha \beta }^{\dagger }|0\rangle }{p^{2}-m^{2}}  \notag \\
&&+\cdots ,
\end{eqnarray}%
where $m_{B^{\ast }}=(5324.75\pm 0.20)~\mathrm{MeV}$ and $m_{D^{\ast
0}}=(2006.85\pm 0.05)~\mathrm{MeV}$ are the masses of the mesons $B^{\ast -}$
and $\overline{D}^{\ast 0}$, whereas their polarization vectors are labeled
by $\varepsilon _{1\mu }$ and $\varepsilon _{2\nu }$, respectively.

The function $\widehat{\Pi }_{\mu \nu \alpha \beta }^{\mathrm{Phys}%
}(p,p^{\prime })$ is found by means of the matrix elements
\begin{eqnarray}
\langle 0|I_{\mu }^{B^{\ast }}|B^{\ast -}(p^{\prime },\varepsilon
_{1})\rangle &=&f_{B^{\ast }}m_{B^{\ast }}\varepsilon _{1\mu }(p^{\prime }),
\notag \\
\langle 0|I_{\nu }^{\overline{D}^{\ast 0}}|\overline{D}^{\ast
0}(q,\varepsilon _{2})\rangle &=&f_{D^{\ast }}m_{D^{\ast 0}}\varepsilon
_{2\nu }(q),  \label{eq:ME2B}
\end{eqnarray}%
with $f_{D^{\ast }}=(252.2\pm 22.66)~\mathrm{MeV}$ and $f_{B^{\ast
}}=(210\pm 6)~\mathrm{MeV}$ being the decay constants of $\overline{D}^{\ast
0}$ and $B^{\ast -}$, respectively. The vertex $\langle B^{\ast -}(p^{\prime
},\varepsilon _{1})\overline{D}^{\ast 0}(q,\varepsilon _{2})|\mathcal{M}_{%
\mathrm{T}}^{\mathrm{b}}(p,\epsilon )\rangle $ is written down in the form
of Eq.\ (\ref{eq:TVV}).

A sum rule for the form factor $g_{3}(q^{2})$ is obtained by employing the
amplitude $\Pi _{3}^{\mathrm{Phys}}(p^{2},p^{\prime 2},q^{2})$ that
corresponds in $\widehat{\Pi }_{\mu \nu \alpha \beta }^{\mathrm{Phys}%
}(p,p^{\prime })$ to the term $\sim p_{\beta }p_{\nu }p_{\mu }^{\prime
}p_{\alpha }^{\prime }$. The correlator $\widehat{\Pi }_{\mu \nu \alpha
\beta }(p,p^{\prime })$ computed using the quark propagators equals to
\begin{eqnarray}
&&\widehat{\Pi }_{\mu \nu \alpha \beta }^{\mathrm{OPE}}(p,p^{\prime })=\frac{%
1}{3}\int d^{4}xd^{4}ye^{ip^{\prime }y}e^{-ipx}\langle \overline{b}b\rangle
\notag \\
&&\times \mathrm{Tr}\left[ \gamma _{\mu }S_{b}^{ia}(y-x)\gamma _{\alpha
}\gamma _{\beta }S_{c}^{aj}(x)\gamma _{\nu }{}S_{u}^{ji}(-y)\right] ,
\label{eq:QCDsideA}
\end{eqnarray}%
where $S_{u}(x)$ is the $u$ quark's propagator \cite{Agaev:2020zad}. We
denote by $\Pi _{3}^{\mathrm{OPE}}(p^{2},p^{\prime 2},q^{2})$ the amplitude
that in $\widehat{\Pi }_{\mu \nu \alpha \beta }^{\mathrm{OPE}}(p,p^{\prime
}) $ corresponds to the same contribution $\sim p_{\beta }p_{\nu }p_{\mu
}^{\prime }p_{\alpha }^{\prime }$.

In what follows, we utilize the relation
\begin{equation}
\langle \overline{b}b\rangle =-\frac{1}{12m_{b}}\langle \frac{\alpha
_{s}G^{2}}{\pi }\rangle  \label{eq:Conden}
\end{equation}%
between the condensates obtained in Ref.\ \cite{Shifman:1978bx}.

The sum rule for the form factor $g_{3}(q^{2})$ reads%
\begin{eqnarray}
&&g_{3}(q^{2})=\frac{2m^{2}(q^{2}-m_{D^{\ast 0}}^{2})}{\Lambda f_{D^{\ast
}}m_{D^{\ast 0}}f_{B^{\ast }}m_{B^{\ast }}}e^{m^{2}/M_{1}^{2}}e^{m_{B^{\ast
}}^{2}/M_{2}^{2}}  \notag \\
&&\times \Pi _{3}(\mathbf{M}^{2},\mathbf{s}_{0},q^{2}).
\end{eqnarray}

For the $B^{\ast -}$ meson channel, we utilize the parameters%
\begin{equation}
M_{2}^{2}\in \lbrack 5.5,6.5]~\mathrm{GeV}^{2},\ s_{0}^{\prime }\in \lbrack
34,35]~\mathrm{GeV}^{2}.  \label{eq:Wind2}
\end{equation}%
To evaluate the coupling $g_{3}$ we use the SR data for $Q^{2}=2-30\ \mathrm{%
GeV}^{2}$ and extrapolating function with parameters $\mathcal{Z}%
_{3}^{0}=0.058~\mathrm{GeV}^{-1}$, $z_{3}^{1}=16.32$, and $z_{3}^{2}=-39.46$%
. The coupling $g_{3}$ is calculated at $q^{2}=m_{D^{\ast 0}}^{2}$
\begin{equation}
g_{3}\equiv \mathcal{Z}_{3}(-m_{D^{\ast 0}}^{2})=(4.4\pm 0.8)\times 10^{-2}\
\mathrm{GeV}^{-1}.  \label{eq:G1}
\end{equation}%
The width of the decay $\mathcal{M}_{\mathrm{T}}^{\mathrm{b}}\rightarrow
B^{\ast -}\overline{D}^{\ast 0}$ is
\begin{equation}
\Gamma \left[ \mathcal{M}_{\mathrm{T}}^{\mathrm{b}}\rightarrow B^{\ast -}%
\overline{D}^{\ast 0}\right] =(11.1\pm 2.9)~\mathrm{MeV}.
\end{equation}

The difference between the decays $\mathcal{M}_{\mathrm{T}}^{\mathrm{b}%
}\rightarrow \overline{B}^{\ast 0}D^{\ast -}$ and $\mathcal{M}_{\mathrm{T}}^{%
\mathrm{b}}\rightarrow B^{\ast -}\overline{D}^{\ast 0}$ is encoded in the
masses of the final-state mesons. With nice accuracy we adopt $\Gamma \left[
\mathcal{M}_{\mathrm{T}}^{\mathrm{b}}\rightarrow \overline{B}^{\ast
0}D^{\ast -}\right] \approx \Gamma \left[ \mathcal{M}_{\mathrm{T}}^{\mathrm{b%
}}\rightarrow B^{\ast -}\overline{D}^{\ast 0}\right] $.


\subsection{Processes $\mathcal{M}_{\mathrm{T}}^{\mathrm{b}}\rightarrow B^{-}%
\overline{D}^{0}$ and $\mathcal{M}_{\mathrm{T}}^{\mathrm{b}}\rightarrow
\overline{B}^{0}D^{-}$}


The channel $\mathcal{M}_{\mathrm{T}}^{\mathrm{b}}\rightarrow B^{-}\overline{%
D}^{0}$ is explored by means of the correlation function
\begin{eqnarray}
\widehat{\Pi }_{\mu \nu }(p,p^{\prime }) &=&i^{2}\int
d^{4}xd^{4}ye^{ip^{\prime }y}e^{-ipx}\langle 0|\mathcal{T}\{I^{B}(y)  \notag
\\
&&\times I^{\overline{D}^{0}}(0)I_{\mu \nu }^{\dagger }(x)\}|0\rangle ,
\end{eqnarray}%
where the currents $I^{\overline{D}^{0}}(x)$ and $I^{B}(x)$ are introduced
by formulas%
\begin{equation}
I^{\overline{D}^{0}}(x)=\overline{c}_{i}(x)i\gamma _{5}u_{i}(x),\text{ }%
I^{B}(x)=\overline{b}_{j}(x)i\gamma _{5}u_{j}(x).
\end{equation}%
To derive the sum rule for the form factor $g_{4}(q^{2})$ that describes the
strong interaction of particles at the vertex $\mathcal{M}_{\mathrm{T}}^{%
\mathrm{b}}B^{-}\overline{D}^{0}$, we have to find the correlators $\widehat{%
\Pi }_{\mu \nu }^{\mathrm{Phys}}(p,p^{\prime })$ and $\widehat{\Pi }_{\mu
\nu }^{\mathrm{OPE}}(p,p^{\prime })$.

We determine $\widehat{\Pi }_{\mu \nu }^{\mathrm{Phys}}(p,p^{\prime })$ by
means of the matrix elements
\begin{equation}
\langle 0|I^{\overline{D}^{0}}|\overline{D}^{0}\rangle =\frac{%
f_{D}m_{D^{0}}^{2}}{m_{c}},\ \langle 0|I^{B}(x)|B^{-}\rangle =\frac{%
f_{B}m_{B}^{2}}{m_{b}}
\end{equation}%
and
\begin{equation}
\langle B^{-}(p^{\prime })\overline{D}^{0}(q)|\mathcal{M}_{\mathrm{T}}^{%
\mathrm{b}}(p,\epsilon )\rangle =g_{4}(q^{2})\epsilon _{\alpha \beta
}^{(\lambda )}(p)p^{\prime \alpha }p^{\prime \beta },
\end{equation}%
where $m_{D^{0}}=(1864.84\pm 0.05)~\mathrm{MeV}$ and $f_{D}=(211.9\pm 1.1)~%
\mathrm{MeV}$ are the mass and decay constant of meson $\overline{D}^{0}$
\cite{PDG:2024,Rosner:2015wva}. The spectroscopic parameters of the meson $%
B^{-}$ are $m_{B}=(5279.41\pm 0.07)~\mathrm{MeV}$ and $f_{B}=206~\mathrm{MeV}
$. As a result, we find
\begin{eqnarray}
&&\widehat{\Pi }_{\mu \nu }^{\mathrm{Phys}}(p,p^{\prime })=\frac{%
g_{4}(q^{2})\Lambda f_{D}m_{D^{0}}^{2}f_{B}m_{B}^{2}}{m_{b}m_{c}\left(
p^{2}-m^{2}\right) \left( p^{\prime 2}-m_{B}^{2}\right) \left(
q^{2}-m_{D^{0}}^{2}\right) }  \notag \\
&&\times \left[ \frac{m^{4}-2m^{2}(m_{B}^{2}+q^{2})+(m_{B}^{2}-q^{2})^{2}}{%
12m^{2}}g_{\mu \nu }\right.  \notag \\
&&\left. +p_{\mu }^{\prime }p_{\nu }^{\prime }+\text{other contributions}%
\right] .
\end{eqnarray}%
For $\widehat{\Pi }_{\mu \nu }^{\mathrm{OPE}}(p,p^{\prime })$, we get%
\begin{eqnarray}
&&\widehat{\Pi }_{\mu \nu }^{\mathrm{OPE}}(p,p^{\prime })=-\frac{1}{3}\int
d^{4}xd^{4}ye^{ip^{\prime }y}e^{-ipx}\langle \overline{b}b\rangle  \notag \\
&&\times \mathrm{Tr}\left[ \gamma _{5}{}S_{b}^{ia}(y-x)\gamma _{\mu }\gamma
_{\nu }S_{c}^{aj}(x)\gamma _{5}S_{u}^{ji}(-y)\right] .  \label{eq:QCDsideB}
\end{eqnarray}%
To obtain the sum rule for $g_{4}(q^{2})$, we employ the amplitudes $\Pi
_{4}^{\mathrm{Phys}}(p^{2},p^{\prime 2},q^{2})$ and $\Pi _{4}^{\mathrm{OPE}%
}(p^{2},p^{\prime 2},q^{2})$ corresponding to structures $p_{\mu }^{\prime
}p_{\nu }^{\prime }$ and find
\begin{eqnarray}
&&g_{4}(q^{2})=\frac{m_{b}m_{c}(q^{2}-m_{D^{0}}^{2})}{\Lambda
f_{D}m_{D^{0}}^{2}f_{B}m_{B}^{2}}e^{m^{2}/M_{1}^{2}}e^{m_{B}^{2}/M_{2}^{2}}
\notag \\
&&\times \Pi _{4}(\mathbf{M}^{2},\mathbf{s}_{0},q^{2}).
\end{eqnarray}

In numerical calculations we have used the parameters
\begin{equation}
M_{2}^{2}\in \lbrack 5.5,6.5]~\mathrm{GeV}^{2},\ s_{0}^{\prime }\in \lbrack
33.5,34.5]~\mathrm{GeV}^{2}.
\end{equation}%
The fit function $\mathcal{Z}_{4}(Q^{2})$ with $\mathcal{Z}_{4}^{0}=0.223~%
\mathrm{GeV}^{-1}$, $\widehat{z}_{4}^{1}=7.397$, and $\widehat{z}%
_{4}^{2}=-7.38$ allows one to estimate the coupling $g_{4}$ which reads
\begin{equation}
g_{4}\equiv \mathcal{Z}_{4}(-m_{D^{0}}^{2})=(2.01\pm 0.36)\times 10^{-1}\
\mathrm{GeV}^{-1}.
\end{equation}%
The function $\mathcal{Z}_{4}(Q^{2})$ and sum rule data are plotted in Fig.\ %
\ref{fig:Fit1}.

The partial width of the decay $\mathcal{M}_{\mathrm{T}}^{\mathrm{b}%
}\rightarrow B^{-}\overline{D}^{0}$ is equal to
\begin{equation}
\Gamma \left[ \mathcal{M}_{\mathrm{T}}^{\mathrm{b}}\rightarrow B^{-}%
\overline{D}^{0}\right] =(13.5\pm 3.4)~\mathrm{MeV}.
\end{equation}%
The width of the process $\mathcal{M}_{\mathrm{T}}^{\mathrm{b}}\rightarrow
\overline{B}^{0}D^{-}$, as it has been explained above, is approximately
equal to $\Gamma \left[ \mathcal{M}_{\mathrm{T}}^{\mathrm{b}}\rightarrow
B^{-}\overline{D}^{0}\right] $.


\subsection{Channels $\mathcal{M}_{\mathrm{T}}^{\mathrm{b}}\rightarrow
\overline{B}_{s}^{\ast 0}D_{s}^{\ast -}$ and $\overline{B}_{s}^{0}D_{s}^{-}$}


Exploration of these processes does not differ considerably from studies of
the modes considered in the previous subsections. Thus, one has to take into
account some substitutions in the correlation functions and in parameters of
new final-state mesons.

In fact, the correlators of the process $\mathcal{M}_{\mathrm{T}}^{\mathrm{b}%
}\rightarrow \overline{B}_{s}^{\ast 0}D_{s}^{\ast -}$ and $\mathcal{M}_{%
\mathrm{T}}^{\mathrm{b}}\rightarrow \overline{B}_{s}^{0}D_{s}^{-}$ can
easily be obtained from Eqs.\ (\ref{eq:QCDsideA}) and (\ref{eq:QCDsideB})
after replacing propagators $S_{u}^{ji}(-y)$ by $S_{s}^{ji}(-y)$. It is
necessary to note that in these calculations we take into account terms $%
\sim m_{s}=(93.5\pm 0.8)~\mathrm{MeV}$, but neglect ones proportional to $%
m_{s}^{2}$. The masses and decay constants of the $\overline{B}_{s}^{(\ast
)0}$ and $D_{s}^{(\ast )-}$ mesons appear as new input parameters: They have
the following numerical values
\begin{eqnarray}
m_{D_{s}} &=&(1969.0\pm 1.4)~\mathrm{MeV},f_{D_{s}}=(249.9\pm 0.5)~\mathrm{%
MeV},  \notag \\
m_{D_{s}^{\ast }} &=&(2112.2\pm 0.4)~\mathrm{MeV},f_{D_{s}^{\ast
}}=(268.8\pm 6.5)~\mathrm{MeV},  \notag \\
&&
\end{eqnarray}%
and
\begin{eqnarray}
m_{B_{s}} &=&(5366.93\pm 0.10)~\mathrm{MeV},\ f_{B_{s}}=234~\mathrm{MeV},
\notag \\
m_{B_{s}^{\ast }} &=&(5415.8\pm 1.5)~\mathrm{MeV},\ \ f_{B_{s}^{\ast }}=221~%
\mathrm{MeV.}
\end{eqnarray}%
These decays are characterized by the strong couplings $g_{5}$ and $g_{6}$
at the vertices $\mathcal{M}_{\mathrm{T}}^{\mathrm{b}}\overline{B}_{s}^{\ast
0}D_{s}^{\ast -}$ and $\mathcal{M}_{\mathrm{T}}^{\mathrm{b}}\overline{B}%
_{s}^{0}D_{s}^{-}$, respectively. The coupling $g_{6}$, for example, is
equal to
\begin{equation}
g_{6}=(1.53\pm 0.28)\times 10^{-1}\ \mathrm{GeV}^{-1}.
\end{equation}%
Relevant SR data and extrapolating function $\mathcal{Z}_{6}(Q^{2})$ are
shown in Fig.\ \ref{fig:Fit1} as well.

The widths of these channels are
\begin{equation}
\Gamma \left[ \mathcal{M}_{\mathrm{T}}^{\mathrm{b}}\rightarrow \overline{B}%
_{s}^{\ast 0}D_{s}^{\ast -}\right] =(8.70\pm 2.34)~\mathrm{MeV},
\end{equation}%
and
\begin{equation}
\Gamma \left[ \mathcal{M}_{\mathrm{T}}^{\mathrm{b}}\rightarrow \overline{B}%
_{s}^{0}D_{s}^{-}\right] =(7.58\pm 2.12)~\mathrm{MeV},
\end{equation}%
respectively. Having used information on decay channels of the hadronic
molecule $\mathcal{M}_{\mathrm{T}}^{\mathrm{b}}$ we estimate its width as%
\begin{equation}
\Gamma \left[ \mathcal{M}_{\mathrm{T}}^{\mathrm{b}}\right] =120_{-12}^{+17}~%
\mathrm{MeV}.
\end{equation}

The result presented above is valid in the case of the first scenario, i.e.,
when the mass $m=15864~\mathrm{MeV}$ makes possible two leading decay modes $%
\mathcal{M}_{\mathrm{T}}^{\mathrm{b}}\rightarrow \Upsilon B_{c}^{\ast -}$and
$\eta _{b}B_{c}^{-}$. In the second case the mass of the molecule $\mathcal{M%
}_{\mathrm{T}}^{\mathrm{b}}$ is $m=15779~\mathrm{MeV}$ which forbids the
channel $\mathcal{M}_{\mathrm{T}}^{\mathrm{b}}\rightarrow \Upsilon
B_{c}^{\ast -}$. The partial widths of numerous decay modes of $\mathcal{M}_{%
\mathrm{T}}^{\mathrm{b}}$ in this scenario are collected in Table \ref%
{tab:MbDecays}, which allow us to estimate the full width of $\mathcal{M}_{%
\mathrm{T}}^{\mathrm{b}}$
\begin{equation}
\Gamma \left[ \mathcal{M}_{\mathrm{T}}^{\mathrm{b}}\right] _{\mathrm{l.l.}%
}=(65\pm 7)~\mathrm{MeV}.
\end{equation}

\begin{table}[tbp]
\begin{tabular}{|c|c|c|c|}
\hline\hline
i & Modes & $g_{i}~(\mathrm{GeV}^{-1})$ & $\Gamma_{i}~(\mathrm{MeV})$ \\
\hline
$2$ & $\eta_b B_c^{-}$ & $22.79 \pm 4.33$ & $6.51 \pm 1.85$ \\
$3$ & $B^{\ast -} \overline{D}^{\ast 0}$ & $(4.28 \pm 0.77)\times 10^{-2}$ & $%
10.12 \pm 2.67$ \\
$4$ & $B^{-}\overline{D}^{0}$ & $(1.92 \pm 0.36)\times 10^{-1}$ & $11.96 \pm
3.37$ \\
$5$ & $\overline{B}_{s}^{ \ast 0}D_{s}^{\ast -}$ & $(3.82 \pm 0.67)\times
10^{-2}$ & $7.94 \pm 2.07$ \\
$6$ & $\overline{B}_{s}^{0}D_{s}^{-}$ & $(1.46 \pm 0.27)\times 10^{-1}$ & $%
6.42 \pm 1.85 $ \\ \hline\hline
\end{tabular}%
\caption{Decay modes of the molecule $\mathcal{M}_{\mathrm{T}}^{\mathrm{b}}$
in the lower mass limit, corresponding strong couplings $g_{i}$
and widths $\Gamma_{i}$.}
\label{tab:MbDecays}
\end{table}


\section{Width of the molecule $\mathcal{M}_{\mathrm{T}}^{\mathrm{c}}$}

\label{sec:Widths3}


In this section we consider the decays of the molecule $\mathcal{M}_{\mathrm{%
T}}^{\mathrm{c}}$. Channels \ $\mathcal{M}_{\mathrm{T}}^{\mathrm{c}%
}\rightarrow J/\psi B_{c}^{\ast +}$, and $\eta _{c}B_{c}^{+}$ dominant modes
of this structure. Annihilation of $c\overline{c}$ quarks give rise to the
processes $\mathcal{M}_{\mathrm{T}}^{\mathrm{c}}\rightarrow B^{\ast
+}D^{\ast 0}$,\ $B^{\ast 0}D^{\ast +}$, $B^{+}D^{0}$,\ $B^{0}D^{+}$, $%
B_{s}^{\ast 0}D_{s}^{\ast +}$, and $B_{s}^{0}D_{s}^{+}$. Because in the
previous sections we have performed detailed analysis of $\mathcal{M}_{%
\mathrm{T}}^{\mathrm{b}}$ molecule's decays, here write down mainly final
predictions for these modes.

As a sample, let us consider the process $\mathcal{M}_{\mathrm{T}}^{\mathrm{c%
}}\rightarrow J/\psi B_{c}^{\ast +}$. The correlator which is required to
find the strong coupling $\widetilde{g}_{1}$ at the vertex $\mathcal{M}_{%
\mathrm{T}}^{\mathrm{c}}J/\psi B_{c}^{\ast +}$ has the form
\begin{eqnarray}
\widetilde{\Pi }_{\mu \nu \alpha \beta }(p,p^{\prime }) &=&i^{2}\int
d^{4}xd^{4}ye^{ip^{\prime }y}e^{-ipx}\langle 0|\mathcal{T}\{I_{\mu
}^{B_{c}^{\ast }}(y)  \notag \\
&&\times I_{\nu }^{J/\psi }(0)\widetilde{I}_{\alpha \beta }^{\dagger
}(x)\}|0\rangle ,
\end{eqnarray}%
with $I_{\mu }^{B_{c}^{\ast }}(x)$ and $I_{\nu }^{J/\psi }(x)$ being the
interpolating currents of the mesons $B_{c}^{\ast +}$ and $J/\psi $
\begin{equation}
I_{\mu }^{B_{c}^{\ast }}(x)=\overline{c}_{i}(x)\gamma _{\mu }b_{i}(x),\
I_{\nu }^{J/\psi }(x)=\overline{c}_{j}(x)\gamma _{\nu }c_{j}(x).
\end{equation}%
The matrix elements used in computations of $\widetilde{\Pi }_{\mu \nu
\alpha \beta }^{\mathrm{Phys}}(p,p^{\prime })$ are
\begin{eqnarray}
&&\langle 0|I_{\mu }^{B_{c}^{\ast }}|B^{\ast +}(p^{\prime },\varepsilon
_{1})\rangle =f_{B_{c}^{\ast }}m_{B_{c}^{\ast }}\varepsilon _{1\mu
}(p^{\prime }),  \notag \\
&&\langle 0|I_{\nu }^{J/\psi }|J/\psi (q,\varepsilon _{2})\rangle =f_{J/\psi
}m_{J/\psi }\varepsilon _{2\nu }(q).
\end{eqnarray}

The correlator $\widetilde{\Pi }_{\mu \nu \alpha \beta }(p,p^{\prime })$ is
given by the expression
\begin{eqnarray}
&&\widetilde{\Pi }_{\mu \nu \alpha \beta }^{\mathrm{OPE}}(p,p^{\prime
})=\int d^{4}xd^{4}ye^{ip^{\prime }y}e^{-ipx}\left\{ \mathrm{Tr}\left[
\gamma _{\mu }S_{c}^{ia}(y-x)\right. \right.  \notag \\
&&\left. \times \gamma _{\alpha }S_{c}^{aj}(x)\gamma _{\nu
}S_{c}^{jb}(-x)\gamma _{\beta }S_{b}^{bi}(x-y)\right] -\mathrm{Tr}\left[
\gamma _{\mu }S_{c}^{ib}(y-x)\right.  \notag \\
&&\left. \left. \times \gamma _{\beta }S_{b}^{bi}(x-y)\right] \mathrm{Tr}%
\left[ \gamma _{\nu }S_{c}^{ja}(-x)\gamma _{\alpha }S_{c}^{aj}(x)\right]
\right\} .
\end{eqnarray}%
The SR for the form factor $\widetilde{g}_{1}(q^{2})$ is derived using the
invariant amplitudes $\widetilde{\Pi }_{1}^{\mathrm{Phys}}(p^{2},p^{\prime
2},q^{2})$ and $\widetilde{\Pi }_{1}^{\mathrm{OPE}}(p^{2},p^{\prime
2},q^{2}) $ which correspond to structures proportional to $p_{\beta }p_{\nu
}p_{\mu }^{\prime }p_{\alpha }^{\prime }$in the correlation functions. After
the Borel transformations and continuum subtractions the amplitude $%
\widetilde{\Pi }_{1}^{\mathrm{OPE}}(p^{2},p^{\prime 2},q^{2})$ has the form
\begin{eqnarray}
&&\widetilde{\Pi }_{1}(\mathbf{M}^{2},\mathbf{s}_{0},q^{2})=%
\int_{(3m_{c}+m_{b})^{2}}^{s_{0}}ds\int_{(m_{b}+m_{c})^{2}}^{s_{0}^{\prime
}}ds^{\prime }\widetilde{\rho }_{1}(s,s^{\prime },q^{2})  \notag \\
&&\times e^{-s/M_{1}^{2}-s^{\prime }/M_{2}^{2}}.
\end{eqnarray}

In numerical computations we employ the masses and decay constants of the
mesons $B_{c}^{\ast +}$ and $J/\psi $. The mass and decay constant of $%
B_{c}^{\ast +}$ have been presented in Sec.\ \ref{sec:Widths1}. As the
spectroscopic parameters of the vector charmonium $J/\psi $ we employ $%
m_{J/\psi }=(3096.900\pm 0.006)~\mathrm{MeV}$, and $\ f_{J/\psi }=(411\pm 7)~%
\mathrm{MeV}$ \cite{PDG:2024,Lakhina:2006vg}.\

Computations of the form factor $\widetilde{g}_{1}(q^{2})$ are performed for
$Q^{2}=2-30~\mathrm{GeV}^{2}$. The Borel and continuum subtraction
parameters $(M_{1}^{2},s_{0})$ in the $\mathcal{M}_{\mathrm{T}}^{\mathrm{c}}$
channel are fixed as in Eq.\ (\ref{eq:Wind1A}), whereas for the $B_{c}^{\ast
+}$ meson channel we use $(M_{2}^{2},s_{0}^{\prime })$ from Eq.\ (\ref%
{eq:Wind2}).

The coupling $\widetilde{g}_{1}$ is obtained at $Q^{2}=-m_{B_{J/\psi }}^{2}$
by means of the extrapolating function $\widetilde{\mathcal{Z}}_{1}(Q^{2})$
which is given by Eq.\ (\ref{eq:FitF}) after replacement $m\rightarrow
\widetilde{m}$. This function has the parameters $\widetilde{\mathcal{Z}}%
_{1}^{0}=0.0676~\mathrm{GeV}^{-1}$, $\widetilde{z}_{1}^{1}=1.403$, and $%
\widetilde{z}_{1}^{2}=0.301$ and leads to the prediction
\begin{equation}
\widetilde{g}_{1}\equiv \widetilde{\mathcal{Z}}_{1}(-m_{B_{J/\psi
}}^{2})=(1.91\pm 0.38)\times 10^{-2}\ \mathrm{GeV}^{-1}.
\end{equation}%
Then the width of the process $\mathcal{M}_{\mathrm{T}}^{\mathrm{c}%
}\rightarrow J/\psi B_{c}^{\ast +}$ is equal to
\begin{equation}
\Gamma \left[ \mathcal{M}_{\mathrm{T}}^{\mathrm{c}}\rightarrow J/\psi
B_{c}^{\ast +}\right] =(19.10\pm 5.48)~\mathrm{MeV}.
\end{equation}

Another dominant and subleading decays of $\mathcal{M}_{\mathrm{T}}^{\mathrm{%
c}}$ are studied in this way as well. Results of relevant calculations are
moved to Table \ref{tab:Channels}, where one can find the strong couplings
and partial widths of these processes. It is worth noting that parameters of
the decays $\mathcal{M}_{\mathrm{T}}^{\mathrm{c}}\rightarrow $\ $B^{\ast
0}D^{\ast +}$ and $\mathcal{M}_{\mathrm{T}}^{\mathrm{c}}\rightarrow
B^{0}D^{+}$\ are not shown, because we take them equal to those of processes
$\mathcal{M}_{\mathrm{T}}^{\mathrm{c}}\rightarrow $\ $B^{\ast +}D^{\ast 0}$
and $B^{+}D^{0}$ and omitted in the table.

Information gained in this section about parameters of different decays
allows one to estimate the full width of molecule $\mathcal{M}_{\mathrm{T}}^{%
\mathrm{c}}$
\begin{equation}
\Gamma \left[ \mathcal{M}_{\mathrm{T}}^{\mathrm{c}}\right] =(71\pm 9)~%
\mathrm{MeV}.
\end{equation}

\begin{table}[tbp]
\begin{tabular}{|c|c|c|c|}
\hline\hline
i & Channels & $\widetilde{g}_{i}~(\mathrm{GeV}^{-1})$ & $\Gamma_{i}~(%
\mathrm{MeV})$ \\ \hline
$1$ & $J/\psi B_c^{\ast +}$ & $(1.91 \pm 0.38)\times 10^{-2}$ & $19.10 \pm
5.48$ \\
$2$ & $\eta_{c} B_c^{+}$ & $(4.61 \pm 0.92)\times 10^{-1}$ & $12.55 \pm 5.45$
\\
$3$ & $B^{\ast +}D^{\ast 0}$ & $(9.27 \pm 1.76)\times 10^{-2}$ & $6.10 \pm
1.67$ \\
$4$ & $B^{+}D^{0}$ & $(6.99 \pm 1.26)\times 10^{-1}$ & $8.87 \pm 2.34$ \\
$5$ & $B_{s}^{\ast 0}D_{s}^{\ast +}$ & $(8.46 \pm 1.69)\times 10^{-2}$ & $%
4.83 \pm 1.37 $ \\
$6$ & $B_{s}^{0}D_{s}^{+}$ & $(5.46 \pm 0.08)\times 10^{-1}$ & $4.65 \pm
1.45 $ \\ \hline\hline
\end{tabular}%
\caption{Processes $\mathcal{M}_{\mathrm{T}}^{\mathrm{c}} \to
B_{(s)}^{(\ast)}D_{(s)}^{(*)}$, related couplings $\widetilde{g}_{i}$ and
widths $\Gamma_{i}$.}
\label{tab:Channels}
\end{table}

\section{Conclusions}

\label{sec:Conc}

The fully heavy hadronic molecules with nonsymmetrical contents are
interesting and relatively new objects for both theoretical and experimental
studies. There are not yet experimental works reported about observation of
such structures. Theoretical studies of such states are also limited by a
few articles.

In the present paper, we explored the hadronic tensor molecules $\mathcal{M}%
_{\mathrm{T}}^{\mathrm{b}}=\Upsilon B_{c}^{\ast -}$ and $\mathcal{M}_{%
\mathrm{T}}^{\mathrm{c}}=J/\psi B_{c}^{\ast +}$ and, for the first time,
calculated their masses and full decay widths. For our investigations we
applied QCD sum rule method which is one of the effective nonperturbative
approaches to evaluate parameters of various hadrons. This method bases on
first principles of QCD and employ universal vacuum expectation values of
different quark-gluon operators. It allows one also to estimate
uncertainties of performed analysis which lacks in most of alternative
methods.

Our predictions $m=(15864\pm 85)~\mathrm{MeV},$ and $\widetilde{m}=(9870\pm
82)~$\textrm{M}$\mathrm{eV}$ for the masses of $\mathcal{M}_{\mathrm{T}}^{%
\mathrm{b}}$ and $\mathcal{M}_{\mathrm{T}}^{\mathrm{c}}$ demonstrate that
they have been estimated with rather high accuracy. These results play a key
role in fixing kinematically allowed decay channels of the molecules. In
turned out that they are strong-interaction unstable states and can
dissociate to constituent mesons or to meson pairs containing all four
initial quarks of $\mathcal{M}_{\mathrm{T}}^{\mathrm{b}}$ and $\mathcal{M}_{%
\mathrm{T}}^{\mathrm{c}}$. These processes are dominant decay modes of the
molecules $\mathcal{M}_{\mathrm{T}}^{\mathrm{b}}$ and $\mathcal{M}_{\mathrm{T%
}}^{\mathrm{c}}$. Another mechanism that permits $\mathcal{M}_{\mathrm{T}}^{%
\mathrm{b}}$ and $\mathcal{M}_{\mathrm{T}}^{\mathrm{c}}$ to transform to
ordinary mesons is annihilation of $b\overline{b}$ and $c\overline{c}$
quarks from their contents.

In the case of the molecule $\mathcal{M}_{\mathrm{T}}^{\mathrm{b}}$
processes $\mathcal{M}_{\mathrm{T}}^{\mathrm{b}}\rightarrow \Upsilon
B_{c}^{\ast -}$, $\eta _{b}B_{c}^{-}$ are dominant ones, whereas six modes $%
\mathcal{M}_{\mathrm{T}}^{\mathrm{b}}\rightarrow B^{(\ast )-}\overline{D}%
^{(\ast )0}$ and $\overline{B}_{(s)}^{(\ast )0}D_{(s)}^{(\ast )-}$ are its
subleading decay channels. In the second scenario the decay $\mathcal{M}_{%
\mathrm{T}}^{\mathrm{b}}\rightarrow \eta _{b}B_{c}^{-}$ is only dominant
channel of the molecule $\mathcal{M}_{\mathrm{T}}^{\mathrm{b}}$. For $%
\mathcal{M}_{\mathrm{T}}^{\mathrm{c}}$ we consider decays to $J/\psi
B_{c}^{\ast +}$ , $\eta _{c}B_{c}^{+}$ , $B^{(\ast )+}D^{(\ast )0}$ and $%
B_{(s)}^{(\ast )0}D_{(s)}^{(\ast )+}$ mesons, where first two channels are
dominant modes, whereas remaining six channels are subleading processes.
Predictions for the decay widths $\Gamma \left[ \mathcal{M}_{\mathrm{T}}^{%
\mathrm{b}}\right] =120_{-12}^{+17}~\mathrm{MeV}$ and $\Gamma \left[
\mathcal{M}_{\mathrm{T}}^{\mathrm{c}}\right] =(71\pm 9)~\mathrm{MeV}$ prove
that these molecules are not rather broad structures. It is worth
emphasizing that these estimations have been made using the central values
of the masses. In the lower limit $15779~\mathrm{MeV}$ of the mass $m$ the
first decay $\mathcal{M}_{\mathrm{T}}^{\mathrm{b}}\rightarrow \Upsilon
B_{c}^{\ast -}$ becomes forbidden for $\mathcal{M}_{\mathrm{T}}^{\mathrm{b}}$
and we get $\Gamma \left[ \mathcal{M}_{\mathrm{T}}^{\mathrm{b}}\right] _{%
\mathrm{l.l.}}=(65\pm 7)~\mathrm{MeV.}$

A similar effect was observed in our previous works \cite%
{Agaev:2025did,Agaev:2025wyf} as well, in which we analyzed features of the
scalar $\eta _{b}B_{c}^{-}$, $\eta _{c}B_{c}^{-}$ and axial-vector $\Upsilon
B_{c}^{-}$, $\eta _{b}B_{c}^{\ast -}$ hadronic molecules. It was proved
there that in lower mass limits the structures $\eta _{b}B_{c}^{-}$ and $%
\Upsilon B_{c}^{-}$ form bound states which are stable against fall-apart
processes, but still can decay to conventional meson pairs through the
annihilation mechanism. These scalar and axial-vector molecules were
explored in Ref.\ \cite{Liu:2024pio}, where the authors employed the
extended local gauge formalism. They found that these structures reside
below the corresponding two-meson thresholds and establish bound states. In
this aspect, our findings agree in part with results of this work.

Note that available theoretical literature for asymmetric fully-heavy
systems is rather limited. In this situation it is instructive to compare
our present results for the tensor molecule with ones obtained in the
diquark-antidiquark picture. As it was emphasized in Sec.\ \ref{sec:Intro},
the tensor tetraquarks were explored in Refs.\ \cite%
{Galkin:2023wox,An:2022qpt,Wu:2024hrv}, in which the authors applied the
relativistic quark model, the constituent and nonrelativistic quark models,
respectively. It was found that the tensor tetraquarks $bb\overline{b}%
\overline{c}$ have the masses $16108~\mathrm{MeV}$, $16149~\mathrm{MeV}$ and
$16554~\mathrm{MeV}$, respectively. In the $cc\overline{c}\overline{b}$
sector of the theory predictions look like $9620~\mathrm{MeV}$, $9731~%
\mathrm{MeV}$, and $10169~\mathrm{MeV}$. It is evident that results of Ref.
\cite{Wu:2024hrv} differ considerably from ones obtained in other
publications\ \cite{Galkin:2023wox,An:2022qpt}. At the same time, last two
works approximately agree with each other in their findings. According to
Ref. \cite{Wu:2024hrv}, for example, there are three tensor resonant states
in the both sectors of the theory.

We have found one ground-level resonant structure in both $bb\overline{b}%
\overline{c}$ and $cc\overline{c}\overline{b}$ cases. Our investigations
lead to the smaller mass in the case of the molecule $\mathcal{M}_{\mathrm{T}%
}^{\mathrm{b}}$, while the mass $\widetilde{m}$ of \ $\mathcal{M}_{\mathrm{T}%
}^{\mathrm{c}}$ exceeds predictions of \ Refs. \cite%
{Galkin:2023wox,An:2022qpt}. In our approach new resonances can be generated
by another tensor molecules with the same contents, for instance, $\Upsilon
(2S)B_{c}^{\ast -}$ and $\psi (2S)B_{c}^{\ast +}$. Investigation of these
structures requires separate detailed analysis which is beyond of the
current article's scope.

Calculations carried out in this article are important to understand
properties of the hadronic tensor molecules $\Upsilon B_{c}^{\ast -}$ and $%
J/\psi B_{c}^{\ast +}$. They provide new and valuable information about
asymmetric fully heavy molecules. It will be interesting to compare our
predictions with results for such hadronic molecules obtained using
alternative methods and, by this way, complete a corresponding theoretical
picture in four-quark mesons' spectroscopy.

\end{document}